\documentclass{aastex63}

\shorttitle{homologous eruption}
\shortauthors{Sahu et al.}

\begin{document}
{

\title{Homologous compact major blowout-eruption solar flares and their production of broad CMEs}

\correspondingauthor{Suraj Sahu}
\email{suraj@prl.res.in}

\author{Suraj Sahu}
\affiliation{Udaipur Solar Observatory, Physical Research Laboratory, Dewali, Badi Road, Udaipur-313001, Rajasthan, India}
\affiliation{Discipline of Physics, Indian Institute of Technology  Gandhinagar, Palaj, Gandhinagar-382355, Gujarat, India}

\author{Bhuwan Joshi}
\affiliation{Udaipur Solar Observatory, Physical Research Laboratory, Dewali, Badi Road, Udaipur-313001, Rajasthan, India}

\author{Alphonse C. Sterling}
\affiliation{NASA/Marshall Space Flight Center, Huntsville, AL 35812, USA}

\author{Prabir K. Mitra}
\affiliation{Udaipur Solar Observatory, Physical Research Laboratory, Dewali, Badi Road, Udaipur-313001, Rajasthan, India}

\affiliation{IGAM/Institute of Physics, University of Graz, Universit{\"a}tsplatz 5, A-8010 Graz, Austria}

\author{Ronald L. Moore}
\affiliation{NASA/Marshall Space Flight Center, Huntsville, AL 35812, USA}
\affiliation{Center for Space Plasma and Aeronomic Research, University of Alabama in Huntsville, Huntsville, AL 35805, USA}

\begin{abstract}

We analyze the formation mechanism of three homologous broad coronal mass ejections (CMEs) resulting from a series of solar blowout-eruption flares with successively increasing intensities (M2.0, M2.6, and X1.0). The flares originated from active region NOAA 12017 during 2014 March 28--29 within an interval of $\approx$24 hr. Coronal magnetic field modeling based on nonlinear-force-free-field extrapolation helps to identify low-lying closed bipolar loops within the flaring region enclosing magnetic flux ropes. We obtain a double flux rope system under closed  bipolar fields for the all the events. The sequential eruption of the flux ropes led to homologous flares, each followed by a CME. Each of the three CMEs formed from the eruptions gradually attain a large angular width, after expanding from the compact eruption-source site. We find these eruptions and CMEs to be consistent with the `magnetic-arch-blowout' scenario: each compact-flare blowout eruption was seated in one foot of a far-reaching magnetic arch, exploded up the encasing leg of the arch, and blew out the arch to make a broad CME.

\end{abstract}

\section{Introduction} \label{sec:intro}
Solar flares are identified as transient, energetic pheneomena occurring in the localized region on the Sun and are responsible for release of huge amount of magnetic energy into interplanetary space. Decades of studies about these phenomena have revealed that solar flares, filament/prominence eruptions, and coronal mass ejections (CMEs) are different observational manifestations of a single physical process occuring in the solar atmosphere and their origin is controlled by local coronal magnetic field topology \citep{Vrsnak2003,Longcope2005,Benz2008,Shibata2011,Green2018}. These various forms of plasma ejections from the Sun play an important role in determining the state of space weather.

Traditionally, solar flares were identified by the manifestation of parallel brightenings (i.e., flare ribbons) on the two sides of a magnetic polarity inversion line (PIL) in an active region of the Sun. Historically, the flare ribbons have been extensively observed in traditional H$\alpha$ observations of the Sun. In the case of an eruptive flare, typically a rising filament-carrying flux rope
stretches the overlying field lines creating a current sheet underneath. Magnetic reconnection sets in at this current sheet and the field lines successively reconnect to form apparently expanding post-flare loops and separating H$\alpha$ ribbons at their footpoints, as the reconnection site achieves successively greater heights in the corona \citep[e.g.,][]{Shibata1999a}. 
To explain the appearance of the flaring loops, two ribbons, and their connection with filament eruptions, a ``standard flare model'' was developed by combining the pioneering works of \citet{Carmichael1964,Sturrock1966,Hirayama1974,Kopp1976}, which is collectively called the `CSHKP model'. The CSHKP model deals with a two-dimensional (2D) configuration of solar flares with a translational symmetry along the reconnecting X-line (the third dimension) \citep{Jing2008}, hence it can also be called a ``standard 2D model'' of solar flares.

In the recent past, there has been significant improvement in our understanding of solar flares owing to sophisticated space-borne satellites and numerical developments, which resulted in the formulation of 3D model of solar flares \citep[e.g.,][]{Aulanier2012}. 
Importantly, this 3D model not only considers the third dimension, but it also includes the strong-to-weak transition of shear from pre-flare magnetic configuration to post-flare loops.
The interpretation of solar flares in the 3D regime is based upon several complex characteristics: coronal sigmoids \citep{Sterling1997,Moore2001,Aulanier2010,Green2011,Joshi2017,Joshi2018,Mitra2018}, systematic HXR footpoint motions along flare ribbons \citep{Fletcher2002,Joshi2009}, 
sheared flare loops \citep{Asai2003,Warren2011}. In addition to these complex magnetic features, some new aspects of the flare ribbons in 3D domain have been identified, such as, photospheric current ribbons \citep{Janvier2014}, three flare ribbons \citep{Wang2014}, circular flare ribbons \citep{Masson2009,Sun2013,Devi2020,Joshi2021,Mitra2021}, J-shaped ribbons \citep{Chandra2009,Schrijver2011,Savcheva2015,NavinJoshi2017} etc. Since the nature of solar flares is intrinsically three dimensional, the energy release processes via magnetic reconnection is supposed to be three dimesional in nature. In view of this idea, there have been several intriguing concepts regarding 3D magnetic reconnection: slipping and slip-running reonnection \citep{Aulanier2006}, null point reconnection \citep{Priest2009,Wang2012,Prasad2020}, interchange reconnection \citep{Fisk2005,Rappazzo2012,Owens2020}, interchange slip-running reconnection \citep{Masson2012}, etc. All these reconnection processes proceed with large-scale restructuring of coronal magnetic fields. Coronal field lines are rooted to the photosphere, which are continuously shuffled by the steady convective motion below the photosphere. This shuffling process interlaces the field lines to generate complex magnetic field topologies, such as: magnetic null points and their associated separatrix surfaces, separator lines, quasi-separatrix layers etc \citep{Demoulin1997,Aulanier2005,Pontin2007a,Jiang2021}.
These intriguing coronal structures act as preferential sites for current accumulation and subsequent magnetic reconnection.
In view of this, in our analysis, we synthesize the multi-wavelength measurements of solar atmospheric layers vis-\`a-vis the topological rearrangement in the coronal magnetic field configuration.


Right from the start of the era of direct soft X-ray (SXR) imaging of the Sun, there has been consensus about the two-element classification of solar flares: eruptive and confined \citep{Pallavicini1977,Svestka1992,Moore2001}. An eruptive flare is one that is accompanied by a CME, whereas a confined flare occurs without a CME. In general, an eruptive flare is characterized by the appearance of two ribbons on both sides of the PIL that spread apart with time as observed in H$\alpha$ images and at other wavelengths. They have large-scale hot post-flare loops observed in SXR and are of long duration (e.g., tens of minutes to a few hours), whereas a confined flare occurs in a relatively much compact region and lasts for a short period (e.g., less than an hour) \citep{Kushwaha2014,Cai2021}. 
Although infrequent, even large X-class flares have also been found to be of confined category \citep{Green2002,Wang2007}. It is worth mentioning that, though the flares and CMEs have a strongly coupled relationship, it is not a cause-effect one \citep{Zhang2001,Temmer2010,Kharayat2021}.

CMEs are designated as `homologous' when they originate from the same location of the active region with similar morphological resemblance in coronagraphic observations \citep{Liu2017}. The cause of occurrence of these homologous CMEs was explored by several authors; \citet{Zhang2002} have stated that repeated flare-CME activities are triggered by the continuous emergence of moving magnetic features in the vicinity of main polarity of the active region. The study of a series of eruptions by \citet{Chertok2004} has revealed that the homology tendency appears to be due to repeated transient perturbation of the global coronal structure, partial eruption, and relatively fast restoration of the same large-scale structures involved in the repeating CME events.
In a magnetohydrodynamic simulation of the development of homologous CMEs by \citet{Chatterjee2013}, the repeated CME activities originate from the repeated formations and partial eruptions of kink unstable flux ropes as a result of continued emergence of twisted flux ropes into a pre-existing coronal potential arcade.

Morphologically, CMEs are complex structures that exhibit a range of shapes and sizes \citep{Schwenn2006,Webb2012}. Recently, it has been recognized that there is a broad class of CMEs, called ``over-and-out'' CMEs, which come from flare-producing magnetic explosions of various sizes and are laterally far offset from the flares. 
A subclass of CMEs of this particular variety was originally identified by \citet{Bemporad2005}, where the authors reported observations of a series of narrow ejections that occurred at the solar limb. These ejections originated from homologous compact flares, whose source was an island of included polarity located just inside the base of a coronal streamer. These ejections resulted in narrow CMEs that moved out along the streamer. It was inferred that each CME was produced by means of the transient inflation or blowing open of an outer loop of the streamer arcade by ejecta, hence they were termed as ``streamer puff'' CMEs. Later \citet{Moore2007} presented new evidence that strengthened the conclusion of \citet{Bemporad2005} and it was inferred that the ``streamer puff'' CMEs are essentially a subgroup of ``over-and-out'' CMEs. For an ``over-and-out'' CME, there would be a laterally-far-offset ejective flare or filament eruption and no discernible flare arcade directly under the CME. Together with the work of \citet{Bemporad2005}, \citet{Moore2007} put forward the concept of `magnetic-arch-blowout' (MAB) to provide a plausible explanation for the production of ``over-and-out'' CMEs. First, a compact magnetic explosion located in a streamer arcade, produces a compact ejective flare. This generates an escaping plasmoid, which becomes the core of the ensuing CME. Second, the source of the explosion, being compact relative to the streamer arcade, should blow out only a short section of the arcade. Observationally, the erupting plasmoid would be laterally deflected by the guiding leg of the streamer arcade and would overpower the arcade near its top, where the arcade field is weaker than its legs. Third, the blowing out of an outer loop of the streamer arcade could result in coronal dimming at the feet of the loop. The lateral extent of the dimming would demarcate the extent of the opened section of the arcade, which participated in the eruption process. Later, \citet{Sterling2011} presented a more generalized concept of MAB scenario, applicable for CMEs that are not produced from streamer regions (see Figure 6 of \citet{Sterling2011}). They investigated two precursor eruptions leading to an X-class flare, where the first precursor was an MAB event. In this case, an initial standard-model eruption of the active region's core field blew out an east-lobe loop of the core region, leading to a CME displaced toward the east of the flaring region. We note that in all the above cases, the basic physical process of eruption remains the same, whereby they differ only in terms of different coronal magnetic environment hosting the compact blowout-eruption flares.

In this paper, we analyze three compact homologous eruptive flares, the eruptions of which result in CMEs of large angular width that resemble the ``over-and-out'' CMEs discussed above. The eruptions originated from the active region NOAA 12017 within a span of $\approx$24 hr during 2014 March 28--29. In this case, each successive eruption is of increasing intensity (i.e., M2.0, M2.6, and X1.0) and result from sequential eruption of low coronal flux ropes lying over the same location of the active region. The third event presented in our analysis (i.e., X1.0 flare) was well observed by a suite of ground- and space-based observatories \citep{Kleint2015,Li2015,Liu2015,Young2015,Woods2018}. The study by \citet{Kleint2015} revealed that a filament eruption was observed above a region of previous flux emergence, which possibly led to a change in magnetic field configuration, causing the X-flare. \citet{Liu2015} discussed a scenario of asymmetric filament eruption due to nonuniform filament confinement and an MHD instability. This disturbed the fan-spine-like field encompassing the filament leading to breakout-type reconnection at the coronal quasi-null region. Subsequently, the filament eruption triggered intense reconnection at the quasi-null, producing a circular flare ribbon. These studies mainly concentrated on a single event (X1.0 flare), which is the strongest among the three major flares. In our study, we analyze the evolution of all three major events. The scope of our study is much broader in that we study the evolution of the three successive major eruptions, all of which originated at the same location of mixed polarity under a dense compact arcade.
Importantly, all the compact magnetic explosions produced an initial perturbation in the system that ultimately resulted in the formation of a broad CME. 
Furthermore, in all three cases, the source region of the CMEs, marked by coronal dimming, exhibit lateral offset from the flaring location -- a typical characteristic of ``over-and-out'' CMEs. 
Although the eruptions of compact flux ropes are progenitors of the CMEs, the actual large-scale structure of the CMEs is linked to the configuration and topological changes in the large-scale magnetic field connecting to magnetic flux far from the flare site.
In Section \ref{sec:observation_data}, we discuss about the observational data sources and techniques. The details of analysis and results are provided in Section \ref{sec:analysis}. The interpretations and conclusions of our analysis are given in Section \ref{sec:discussion}.

\begin{figure}[htp!]
\hspace{1.2cm}
\includegraphics[scale=1.5]{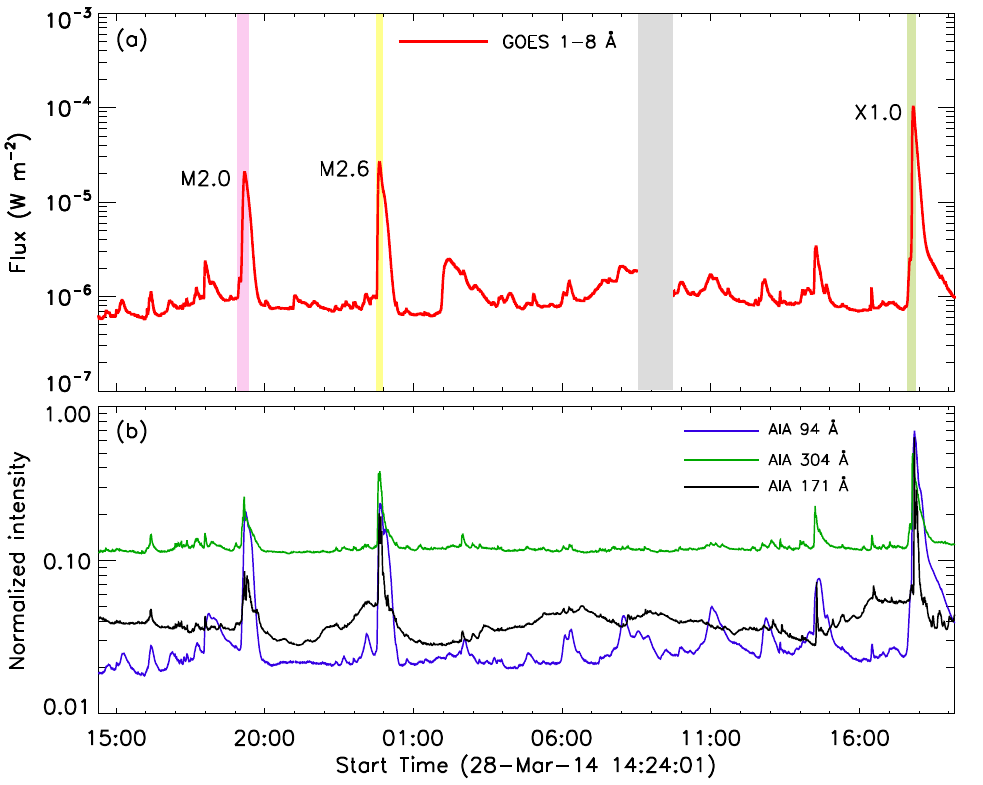}
\caption{Panel (a): GOES soft X-ray flux in 1--8 \AA\ channel indicating the three flare events of intensities M2.0, M2.6, and X1.0. The shaded regions in purple, yellow, and olive green represent the flare durations according to the GOES flare catalog. The gray shaded region indicates a period over which GOES data were unavailable. Panel (b): Normalized intensity light curves of AIA passbands 94 \AA\ [log(T)=6.8)], 304 \AA\ [log(T)=4.7)], and 171 \AA\ [log(T)=5.7)] multiplied by factors of 0.7, 0.5, and 1.2, respectively, for clear visualization. The light curves denote the variation of intensity over the AR under analysis.}
\label{fig:goes}
\end{figure}

\section{Observations and data sources}
\label{sec:observation_data}

We have extensively used data taken from Atmospheric Imaging Assembly \citep[AIA;][]{Lemen2012} on board the Solar Dynamics Observatory \citep[SDO;][]{Pesnell2012} for extreme ultraviolet (EUV) imaging. AIA observes the full disk of the Sun and produces images in seven EUV (94 \AA, 131 \AA, 171 \AA, 193 \AA, 211 \AA, 304 \AA, and 335 \AA) and two ultraviolet (UV) (1600 \AA\ and 1700 \AA) passbands. The images produced by AIA have cadence of 12 s and 24 s for EUV and UV filters, respectively, and a pixel resolution of 0$\farcs$6, while the instrumental resolution of AIA is 1$\farcs$5}.

We have used line-of-sight (LOS) magnetic field, intensity, and vector magnetic field measurements at the solar photosphere recorded by the Heliosesmic and Magnetic Imager \citep[HMI;][]{Schou2012} on board SDO. The cadence of LOS magnetic field and intensity measurement is 45 s, whereas the vector magnetic field has cadence of 720 s. The data obtained from HMI have resolution of 0$\farcs$5 pixel$^{-1}$. We employ the SolarSoftWare \citep[SSW\footnote{\url{https://www.lmsal.com/solarsoft/}};][]{Freeland2012} routine \textit{hmi\_prep.pro}, which converts the resolution to 0$\farcs$6 pixel$^{-1}$ to compare the HMI data with the AIA data. The instrumental resolution of HMI is 1$\farcs$0.

The CMEs of the events under analysis were observed by the C2 coronagraph of the Large Angle and Spectrometric Coronagraph \citep[LASCO;][]{Brueckner1995} on board the Solar and Heliospheric Observatory \citep[SOHO;][]{Domingo1995}. The C2 coronagraph observes the solar corona in white light images with a field-of-view (FOV) of 1.5--6 R$_\odot$.

To understand the small-scale coronal magnetic configurations in the pre-flare stages, we employ the Nonlinear-force-free-field (NLFFF) method developed by \citet{Wiegelmann2008}, which utilizes the `optimization approach' \citep{Wheatland2000, Wiegelmann2004}. We use HMI.sharp$\textunderscore$cea$\textunderscore$720s series vector magnetograms for the photospheric boundary condition as inputs of the extrapolation. Our extrapolation volume extends up to 385, 316, and 252 pixels in x, y, and z directions, respectively, taking the photosphere as the x-y plane. These correspond to a physical volume of (280 $\times$ 229 $\times$ 183) Mm$^{3}$ above the active region (AR) under consideration. For visualization of the extrapolated magnetic field lines in 3D, we use the Visualization and Analysis Platform for Ocean, Atmosphere, and Solar Researchers \citep[VAPOR\footnote{\url{https://www.vapor.ucar.edu/}};][]{Clyne2007} software.

The large-scale coronal magnetic field lines surrounding the activity site are determined using the Potential-Field-Source-Surface (PFSS) extrapolation method \citep{Schrijver2003}. PFSS is an IDL based algorithm that is available in the SSW package.

\begin{deluxetable}{ccccccc}[htp!]
\tablecolumns{6}
\tablewidth{6pt}

 \tablecaption{Summary of the flares in AR NOAA 12017 during 2014 March 28 to 29 \label{tab:summary}}

 \tablehead{
 \colhead{\hspace{-0.1cm} Event} & \colhead{\hspace{0.5cm}Flare} & \colhead{\hspace{0.5cm}Date} & \colhead{} & \colhead{Time (UT)} & \colhead{} & \colhead{\hspace{0.5cm}Heliographic}\\
\colhead{\vspace{-0.1cm}number} & \colhead{\hspace{0.5cm}class} & \colhead{} & \colhead{\hspace{0.5cm}Start} & \colhead{\hspace{0.5cm}Peak} & \colhead{\hspace{0.5cm}End} & \colhead{\hspace{0.5cm}coordinate}\
}

 \startdata 
 I & \hspace{0.5cm}M2.0 & \hspace{0.5cm}2014 March 28 & \hspace{0.5cm}19:04 & \hspace{0.5cm}19:18 & \hspace{0.5cm}19:27 & \hspace{0.5cm}N11W21\\
II & \hspace{0.5cm}M2.6 & \hspace{0.5cm}2014 March 28 & \hspace{0.5cm}23:44 & \hspace{0.5cm}23:51 & \hspace{0.5cm}23:58 & \hspace{0.5cm}N11W23\\
III & \hspace{0.5cm}X1.0 & \hspace{0.5cm}2014 March 29 & \hspace{0.5cm}17:35 & \hspace{0.5cm}17:48 & \hspace{0.5cm}17:54 & \hspace{0.5cm}N11W32\\
 \enddata

\end{deluxetable}

\section{Analysis and results}
\label{sec:analysis}

\subsection{Relationship between Coronal mass ejections and their source region}
\label{sec:CME_and_source}

In Figure \ref{fig:goes}(a), we provide the GOES light curve in 1--8 \AA\ channel from $\approx$14:30 UT on 2014 March 28 to $\approx$19:00 UT on 2014 March 29. The light curve clearly indicates the occurrence of three large flares of class M2.0, M2.6, and X1.0 (the flare intervals are marked by vertical shaded regions in different colors). All these flares were of the eruptive category and produced spectacular, large-scale structures of CMEs observed in the white light coronagraphic images from LASCO\footnote{\url{https://cdaw.gsfc.nasa.gov/CME_list/UNIVERSAL/2014_03/univ2014_03.html}}. We find that the eruptive flares show successively increasing intensities (viz., M2.0, M2.6, and X1.0). We also note that there had been no significant flaring activities (of class $>$M) at least one day before and after the events under analysis. These events occurred in AR NOAA 12017, which presented a $\beta\gamma$ type photosphetric magnetic configuration during the reported activities. This AR is typical in the sense that it exhibited multiple flaring episodes within a short span of $\approx$two days during 2014 March 28--29, and it has been the subject of several studies. \citet{Yang2016} investigated the magnetic field of the AR during the aforementioned period and found the presence of a magnetic flux rope (MFR) in an NLFFF extrapolation, which was prone to kink instability. Furthermore, the closed quasi-separatrix layer structure surrounding the MFR became smaller as a consequence of the eruption. \citet{Chintzoglou2019} revealed that AR NOAA 12017 hosted a `collisional PIL', which developed owing to the collision between two emerging flux tubes nested within the AR. Also, during the entire evolution over the visible solar disk, the AR showed significant cancellation (up to 40\%) of the unsigned magnetic flux of the smallest emerging bipolar magnetic region.

In Figure \ref{fig:goes}(b), we plot the EUV light curves based on AIA imaging observations in 94 \AA, 171 \AA, and 304 \AA\ channels. The AIA light curves are obtained from intensity variation over the whole AR NOAA 12017 (marked by a red rectangle in Figure \ref{fig:pfss1}(b)). We note that the EUV 94 \AA\ light curve resembles the GOES 1--8 \AA\ light curve fairly well.
A summary of the flaring events is given in Table \ref{tab:summary}, which is based on GOES flare catalog\footnote{\url{https://www.ngdc.noaa.gov/stp/space-weather/solar-data/solar-features/solar-flares/x-rays/goes/xrs/goes-xrs-report_2014.txt}}.

\begin{figure}
\hspace{2.25cm}
\includegraphics[scale=0.5]{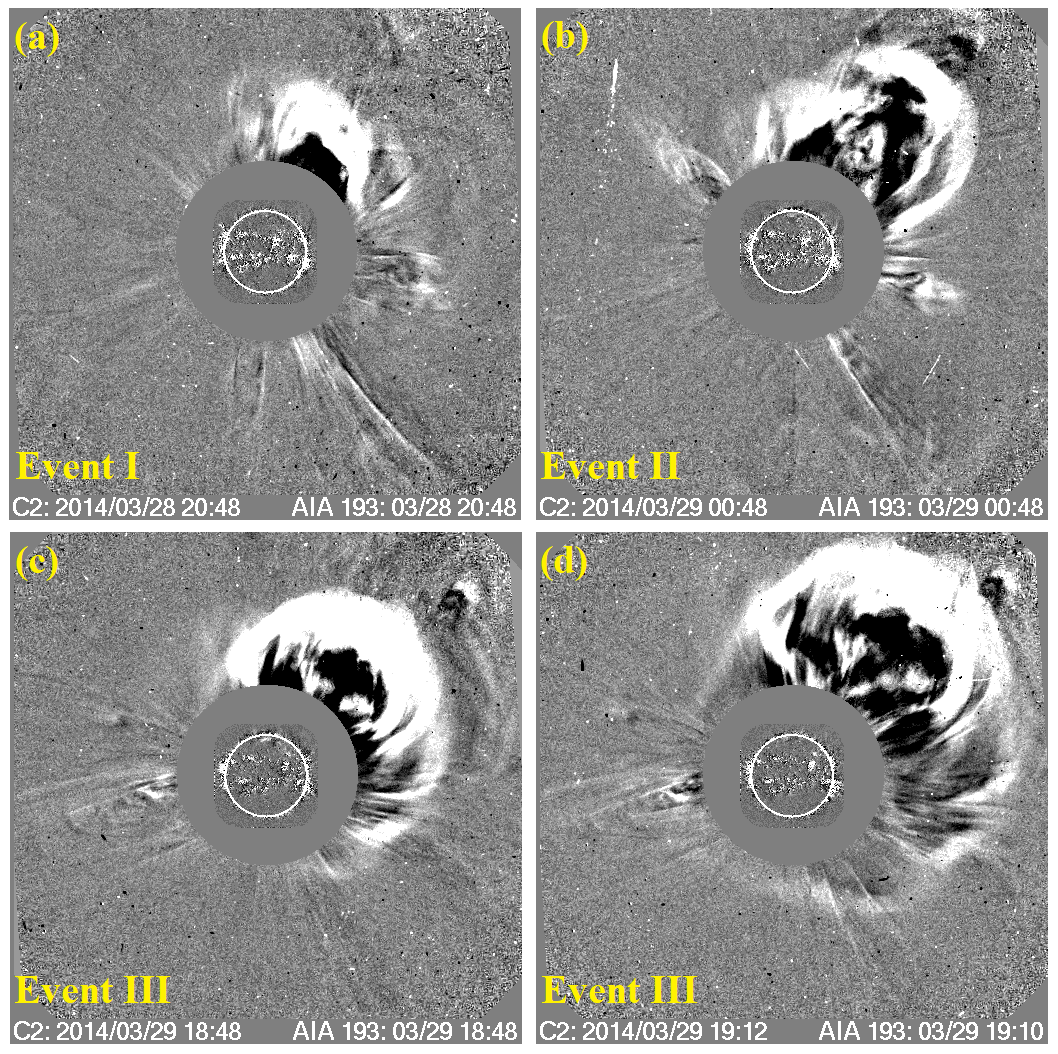}
\caption{Depiction of the wide CMEs formed due to the large-scale eruptions associated with the events under analysis. Panels (a) and (b) show the CMEs associated with the events I and II and the CME produced in the aftermath of event III is shown in panels (c) and (d). In all panels, the CME images are from the LASCO C2 coronagraph and the coronagraph occulter is overplotted by an AIA 193 \AA\ image.}
\label{fig:CME}
\end{figure}

\begin{figure}[htp!]
\hspace{2.35cm}
\includegraphics[scale=0.3]{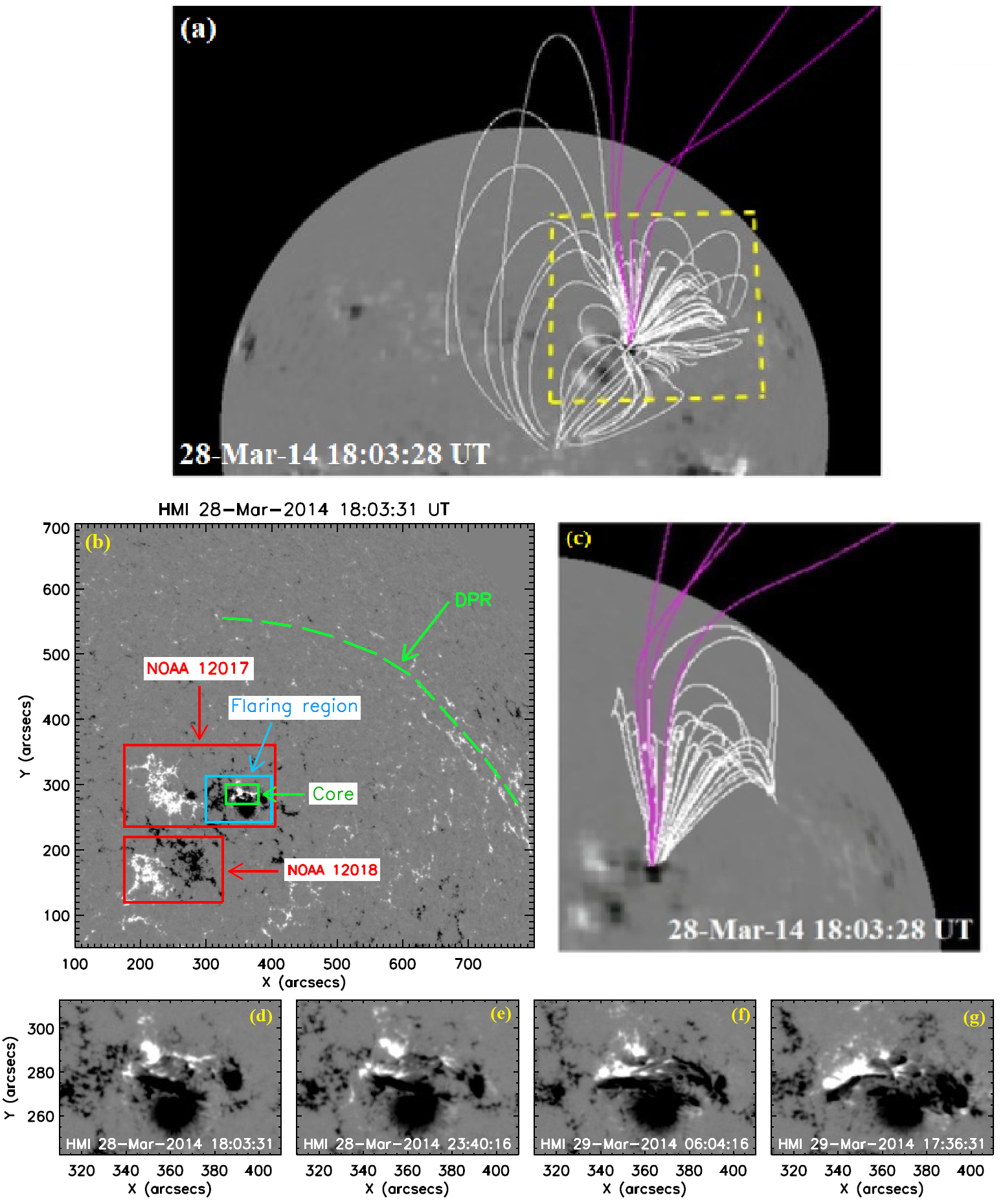}
\caption{Panel (a): The large-scale field surrounding the active region from which the eruptions occur. The white and pink field lines denote the closed and open field lines, respectively, overlaid onto LOS magnetogram. The yellow dashed box is enlarged in panels (b)--(c). In panel (b), we show the photospheric LOS magnetogram. The ARs NOAA 12017 and 12018 are shown by the red rectangles. We note that NOAA 12017, the AR of our interest, displays approximate spatial extension of 230$\arcsec\times$125$\arcsec$. We mark the flaring region within this AR by a sky blue box. Inside this, we mark the `core' location of mixed polarity magnetic field by a green box. The spatial extension of the core is $\approx$50$\arcsec\times$30$\arcsec$, and it is the source of all the compact blowout-eruption flares. Notably, the core region is much smaller ($\approx$19 times) than the size of the AR. We show the distant positive polarity region (DPR) by a green dashed line. In panel (c), we show the large-scale connectivity between the negative polarity of the flaring region and the DPR by white field lines (cf. panels (b) and (c)) before the time of event I. The pink field lines are open field lines originating from the flaring region. In panels (d)--(g), we show the evolution of the flaring region. In panels (d) and (e), we show the morphology of this region before events I and II, respectively, whereas, the magnetic configuration before event III is shown in panels (f) and (g).}
\label{fig:pfss1}
\end{figure}

In Figure \ref{fig:CME}, we provide LASCO C2 images of three CMEs (marked as Event I, II, and III). The CMEs associated with events I and II are non halo (with angular sizes of 103$^{\circ}$ and 111$^{\circ}$, respectively), whereas the CME associated with event III is a halo CME. We note that the linear speed of the CMEs increases gradually from event I to event III, with values of 420, 503, and 528 km s$^{-1}$, respectively. Several parameters of these CMEs are listed in Table \ref{tab:CME}.

\begin{deluxetable}{cccccc}[htp!]
\tablecolumns{6}
\tablewidth{6pt}

 \tablecaption{Some parameters of the CMEs made by the events under investigation \label{tab:CME}}

 \tablehead{
 \colhead{CME} & \colhead{\hspace{0.7cm}Speed (km s$^{-1}$)} & \colhead{} & \colhead{Angular width} & \colhead{Mass} & \colhead{Kintic energy}\\
\colhead{} & \colhead{\hspace{-1cm}V$_{L}$} & \colhead{\hspace{-1.6cm}V$_{S}$} & \colhead{(degree)} & \colhead{($\times$10$^{15}$ gram)} & \colhead{($\times$10$^{30}$ erg)}\
}

 \startdata 
 CME 1 & \hspace{-1cm}420 &  \hspace{-1.6cm}464 & 103 & 2.6 & 2.3\\
 CME 2 & \hspace{-1cm}503 &  \hspace{-1.6cm}327 & 111 & 2.3 & 2.9\\
 CME 3 &  \hspace{-1cm}528 &  \hspace{-1.6cm}505 & 360 & 5.0 & 7.0\\
 \enddata
\tablecomments{CME 1, CME 2, and CME 3 are made by events I, II, and III, respectively. V$_{L}$ and V$_{S}$ are linear speed and second order speed at 20 solar radii, respectively. The  various parameters of CMEs are obtained from the SOHO/LASCO CME catalog\footnote{\url{https://cdaw.gsfc.nasa.gov/CME_list/UNIVERSAL/2014_03/univ2014_03.html}}.}
\end{deluxetable}

\begin{figure}[htp!]
\hspace{1.2cm}
\includegraphics[scale=0.6]{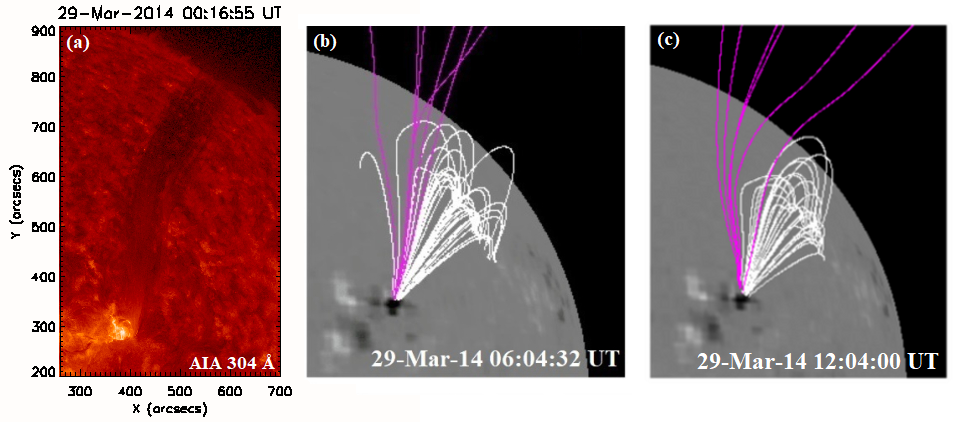}
\caption{In panel (a), we show a surge consisting of cool material expelled from the flaring region, observed to occur after event II. Panels (b) and (c) show the large-scale connectivity, revealed by PFSS model extrapolation, between the flaring region and the DPR at instances shortly after event II and shortly before event III, respectively. The surge nearly follows the open pink field lines (cf. panels (a), (b), and (c)). The large-scale connectivity between the flaring region and the DPR remains unchanged before event I, after event II, and before event III (cf. panels (b) and (c) with Figure \ref{fig:pfss1}(c)).}
\label{fig:pfss2}
\end{figure}

The large-scale coronal connectivities are thought to play a major role in the development of broad CME structures. In order to visualize large-scale coronal magnetic field lines, we carry out PFSS extrapolation for a few representative instances (see Figures \ref{fig:pfss1}(a), \ref{fig:pfss1}(c), and \ref{fig:pfss2}(b)--(c)). In Figure \ref{fig:pfss1}(a), we show the extrapolated field lines in and around the active region before event I. To show the detailed magnetic structure on the photosphere, the region marked within the yellow dashed box in Figure \ref{fig:pfss1}(a) has been shown in Figures \ref{fig:pfss1}(b)--(c).
In Figure \ref{fig:pfss1}(b), ARs NOAA 12017 and 12018 are marked by red rectangles. NOAA 12017 is the AR of our interest. The flaring region lies in the leading part of the AR, which is marked by a sky blue box. In all three cases, the eruption begins with a compact blowout eruption of a flux rope from a small region of mixed polarity within the flaring region. We term this small region as the `core', and mark it by a green box (see also Figure \ref{fig:flux_ropes}). Notably, the size (i.e., area under green box) of the core region is significantly smaller ($\approx$19 times) than the extent of the AR (i.e., the area under the red rectangle denoting AR NOAA 12017). 
A careful examination of the magnetogram reveals the presence of extended and dispersed magnetic flux of positive polarity, located toward the north-west of the flaring site, substantially away from the AR boundaries. This distant positive-polarity region (abbreviated as DPR) extends like an arc, which we manually mark by a green dashed line in Figure \ref{fig:pfss1}(b). 
Importantly, we note clear magnetic connectivities between the negative polarity of the flaring region and the DPR, which we denote as white field lines in Figure \ref{fig:pfss1}(c), adjacent to open field lines (shown in pink), originating from the flaring region. The DPR acts as a proxy for the remote foot points of large-scale coronal field lines (i.e., a magnetic arch), which is involved in the formation of large-scale CMEs.
To explore the fine details of the flaring region, we show zoomed images of this region in Figures \ref{fig:pfss1}(d)--(g). Panels (d) and (e) represent the magnetic configuration before events I and II, whereas panels (f) and (g) represent the magnetic structure of the flaring region before event III. Comparison of panels (d)--(g) of Figure \ref{fig:pfss1} clearly reveals substantial small-scale changes (i.e., various epochs of emergence and cancellation) in photospheric magnetic field of the flaring region over the time period spanning the three events.

Each of the eruptive flares is followed by collimated surges of cool material from the flaring region. In Figure \ref{fig:pfss2}(a), we show a representative image of the surge from the flaring region observed after event II. We choose this particular observation because of its clear visibility. In Figures \ref{fig:pfss2}(b)--(c), we show the large-scale coronal field lines, demonstrated by PFSS extrapolation, connecting the flaring region with the DPR (shown by white field lines). The epochs in these panels denote instances a few hours after event II and a few hours before event III, respectively. The pink lines represent open field lines emanating from the flaring region. We note that the large-scale coronal magnetic field configuration remains unchanged during the course of the events (cf. Figures \ref{fig:pfss1}(c) and \ref{fig:pfss2}(b)--(c)). Interestingly, the surge nearly follows the open field lines (cf. Figures \ref{fig:pfss2}(a)--(c)).

\begin{figure}[htp!]
\centering
\includegraphics[scale=1.1]{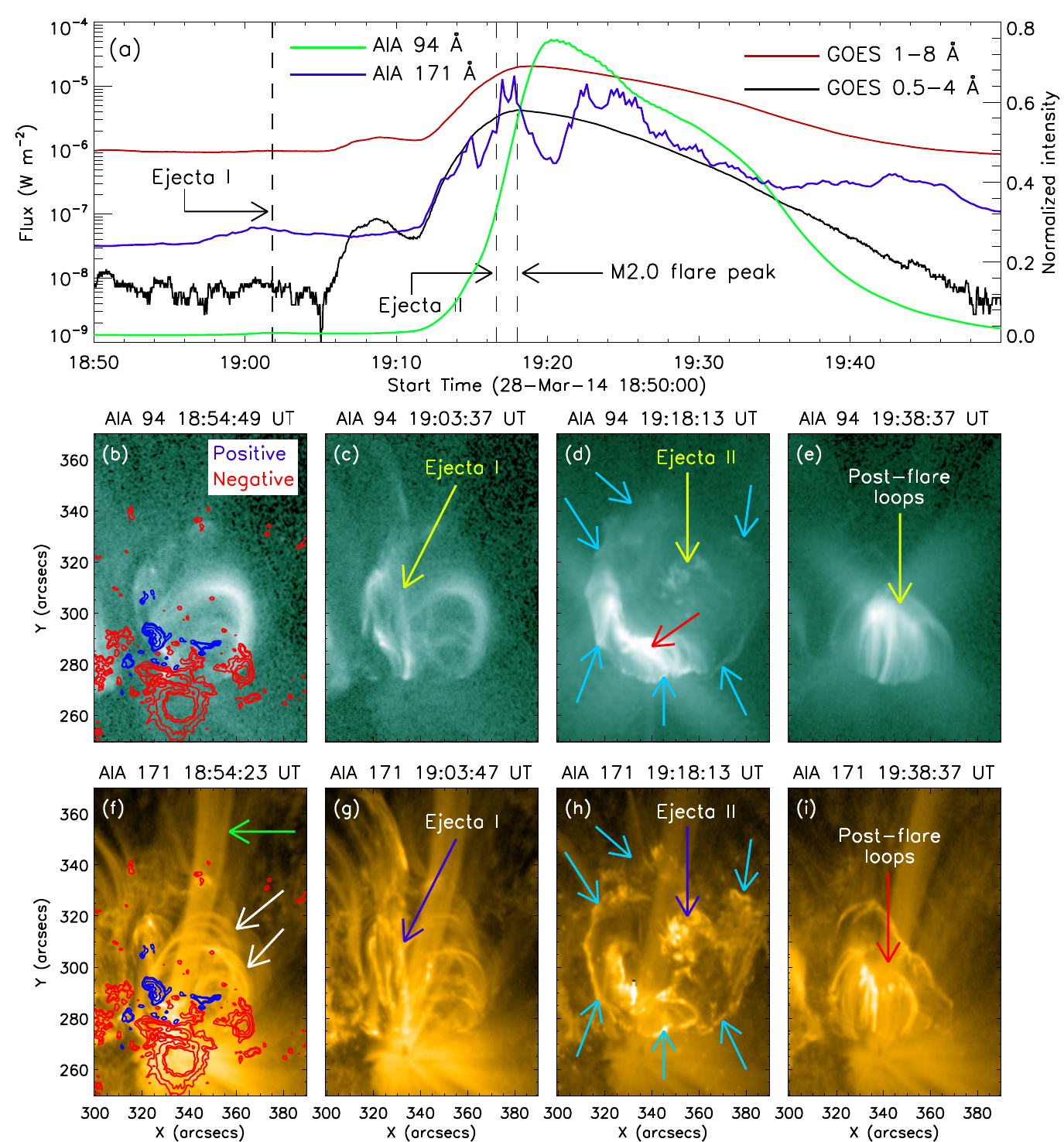}
\caption{Panel (a): The GOES lightcurves in 1--8 \AA\ and 0.5--4 \AA\ channels and the intensity curves for event I, obtained from the flaring region recorded in AIA 94 \AA\ and 171 \AA\ channels. The light curves span from 18:50 UT to 19:50 UT on 2014 March 28 showing the epochs of the first event, which had the M2.0 flare and two associated ejecta. Panels (b)--(e): Evolution of the eruption in the AIA 94 \AA\ channel. In panel (b), we overlay the LOS magnetic contours with red and blue, representing negative and positive magnetic polarities, respectively. Panels (c) and (d) indicate the ejecta I and II, respectively. In panels (d) and (h), we demarcate a wide circular ribbon structure north of the core region by sky blue arrows. The compact post-flare loops resulted from the standard flare reconnection between the legs of the field lines stretched by the erupting flux rope are indicated by a red arrow in panel (d). In panel (e), we show the growing post-flare loop arcades. Panels (f)--(i) show the flare evolution in the AIA 171 \AA\ channel. The same magnetic contours as in panel (b), are also shown in panel (f). The magnetic contour levels are set as $\pm$[200, 400, 800, 1000, 2000]G. In panel (f), the two white arrows indicate small-scale connectivities within the flaring region, whereas the green arrow shows several structures that extend outward. Ejecta I, ejecta II, and the post-flare arcades are indicated in panels (g), (h), and (i), respectively. The peak of the flare occurs just after the eruption of ejecta II (see panel (a)).}
\label{fig:AIA_M2.0}
\end{figure}

\begin{figure}[htp!]
\hspace{1.5cm}
\includegraphics[scale=1.1]{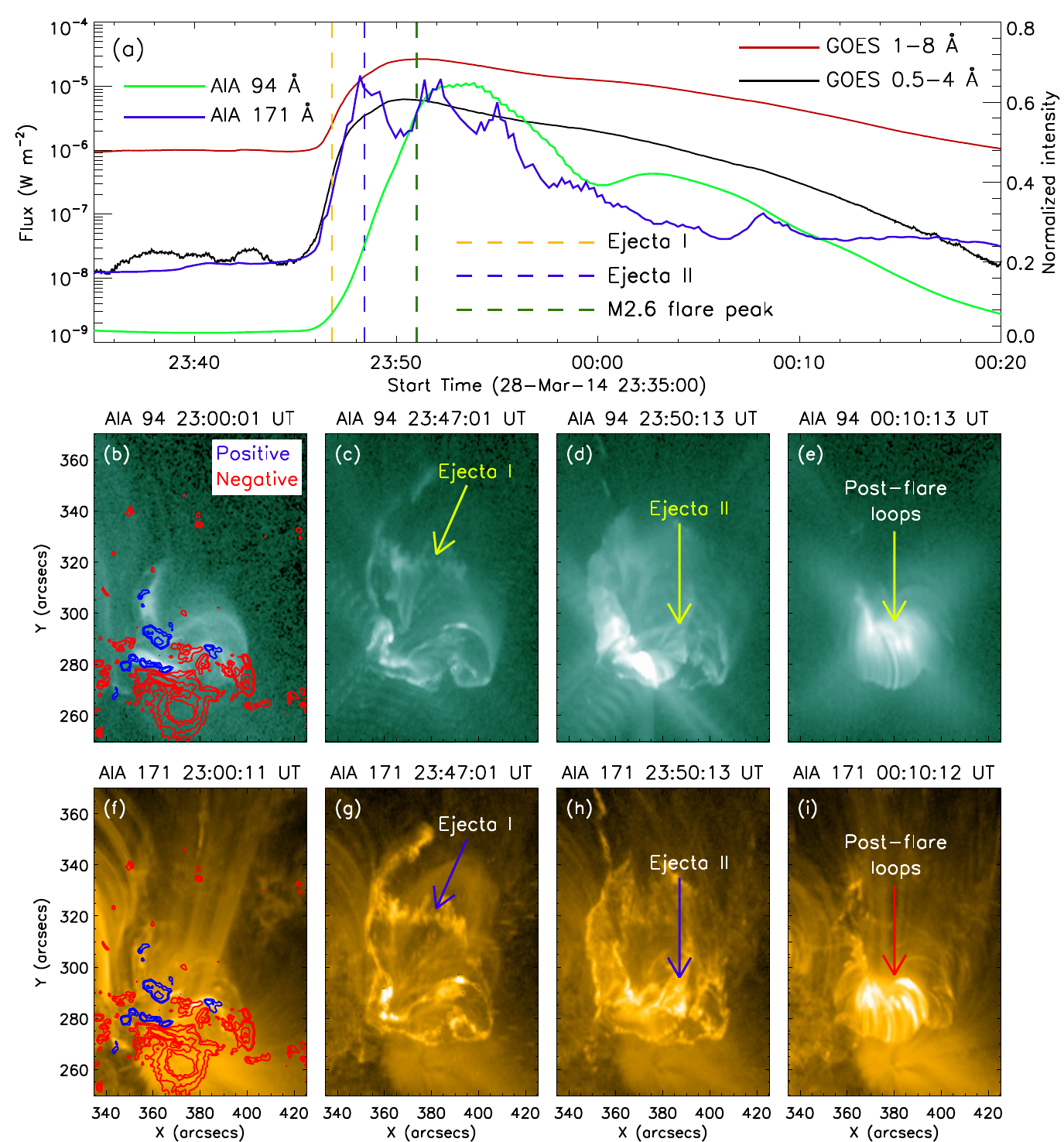}
\caption{Panel (a): The GOES light curves in 1--8 and 0.5--4 \AA\ channels along with the intensity curves obatained from the flaring region in AIA 94 \AA\ and 171 \AA\ channels for event II. The time range of the light curves spans between 23:35 UT on 2014 March 28 and 00:20 UT on March 29. The flare peak and the two ejecta are indicated by dashed lines in different colors. Panels (b)--(e): Evolution of the flare shown in the AIA 94 \AA\ channel. The LOS magnetogram is overplotted as contours on the 94 \AA\ image in panel (a), with colors as denoted in Figure. Ejecta I and II are shown in panels (c) and (d), respectively. The post-flare loops are shown in panel (e). In panels (f)--(i), we show the flare evolution in AIA 171 \AA\ observations. The LOS magnetogram is overplotted on the 171 \AA\ image in panel (f). The magnetic contour levels are $\pm$[200, 400, 800, 1000, 2000]G in all the panels. Panels (g) and (h) show the ejecta I and II, respectively. The formation of bright post-flare loops are shown in panel (i).}
\label{fig:AIA_M2.6}
\end{figure}

\subsection{ Trio of blowout-eruption flares and associated magnetic environment} 
\label{sec:multi-wavelength_results}

\begin{figure}[htp!]
\hspace{1.5cm}
\includegraphics[scale=1.1]{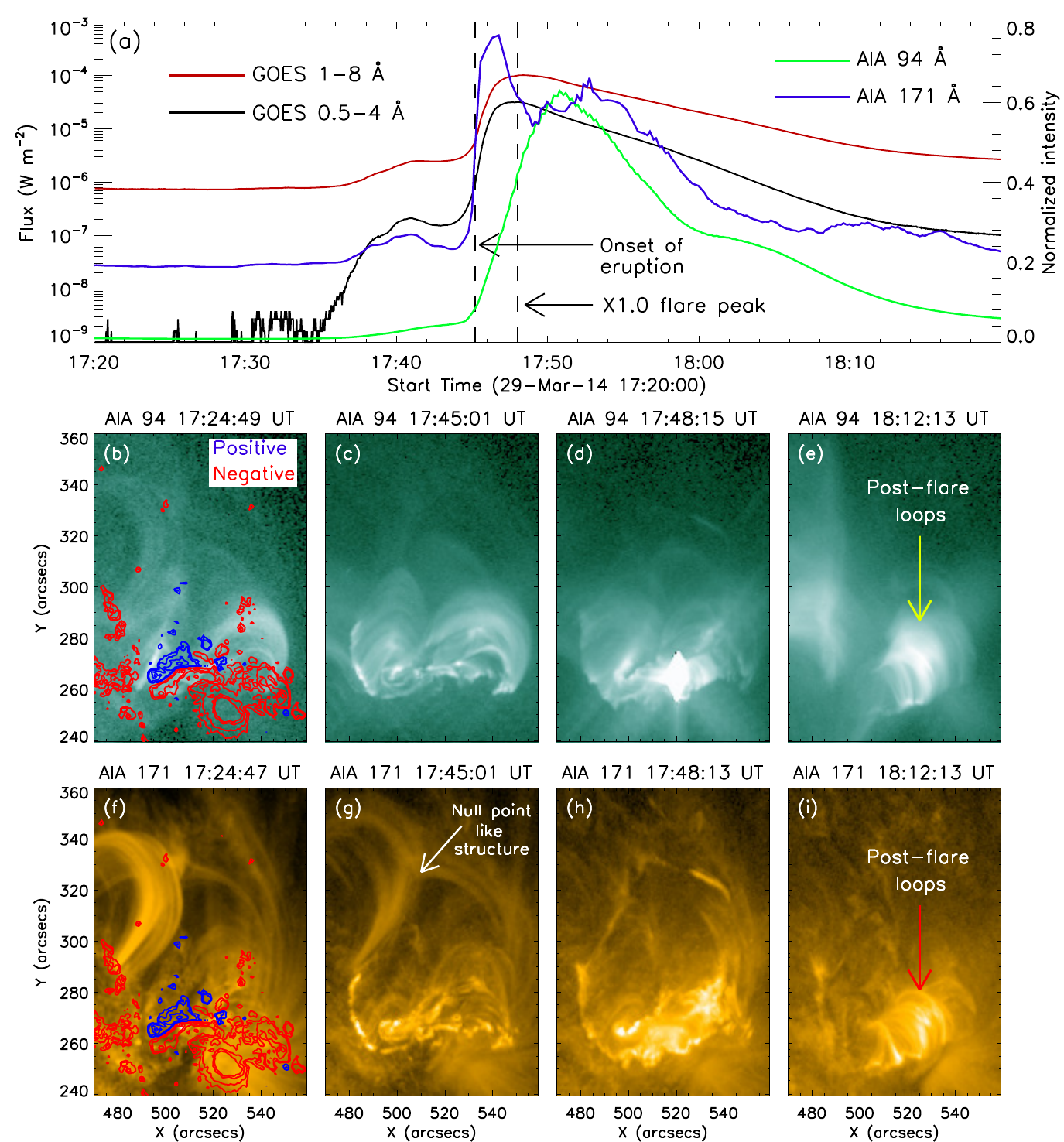}
\caption{In panel (a), we plot the GOES light curves in 1--8 \AA\ and 0.5--4 \AA\ channels for the case of event III, along with the intensity curves from the AIA 94 \AA\ and 171 \AA\ channels recorded from the flaring region. The time interval chosen for this panel runs from 17:20 UT to 18:20 UT on 2014 March 29. We indicate the onset of eruption and the flare peak by dashed lines. In this case, we observe a single eruption from the core region, unlike events I and II where we observed two successive eruptions denoted as ejecta I and II in Figures \ref{fig:AIA_M2.0} and \ref{fig:AIA_M2.6}. Panels (b)--(e) show the evolution of the flare in AIA 94 \AA\ images. In panel (b), we overplot the magnetic contours onto the 94 \AA\ image. In panel (e), we show the post-flare loop arcades. In panels (f)--(i), we show the flare evolutuion in the AIA 171 \AA\ channel. The magnetic contours drawn in panel (f) are the same as in panel (b). The contour levels are set as $\pm$[200, 400, 800, 1000, 2000]G. We observe an inverted Y-shaped null-point-like structure in the pre-eruptive stage, indicated in panel (g). The formation of post-flare loop arcades is shown in panel (i).}
\label{fig:AIA_X1.0}
\end{figure}

The temporal and morphological analysis of the eruptive flares are presented in Figures \ref{fig:AIA_M2.0}, \ref{fig:AIA_M2.6}, and \ref{fig:AIA_X1.0}. Panel (a) in these figures show the GOES light curves of the events in the 1--8 \AA\ and 0.5--4 \AA\ channels, along with the AIA light curves in 94 \AA\ and 171 \AA, while panels (b)--(i) provide a few representative AIA images. For imaging analysis, we examine AIA observations taken in the 94 \AA\ and 171 \AA\ channels. The AIA 94 \AA\ [log(T)=6.8)] channel is apt for imaging the flaring coronal structures while AIA 171 \AA\ [log(T)=5.7)] channel is useful to infer the low temperature structures formed in the corona and transition region.  
The selected FOV of the AIA images encompasses the flaring region (see Figures \ref{fig:pfss1}(d)--(g)) and surrounding regions into which the eruption evolves.

Figure \ref{fig:AIA_M2.0} reveals several temporal and spatial aspects of the first (M2.0 flare) event. 
A comparison of LOS photopheric magnetic flux with the EUV images (see panels (b) and (f)) during the pre-flare stage reveals small-scale connectivities (shown by white arrows in panel (f)) between opposite magnetic polarities within the core region. In the 171 \AA\ image in panel (f), we note several structures that extend outward (shown by green arrow), suggestive of either large-scale or quasi-open field lines. The presence of open field lines is well supported by the global PFSS extrapolation presented in Figures \ref{fig:pfss1} and \ref{fig:pfss2} (see also Section \ref{sec:CME_and_source}). The sequence of AIA images reveals two stages of eruptions, which we term as ejecta I and II. Ejecta I originates from the eastern part of the core at $\approx$19:03 UT (marked as `ejecta I' in panels (c) and (g)). Ejecta II starts at $\approx$19:18 UT from the western part of the core (marked as `ejecta II' in panels (d) and (h)). The onset times of the two ejecta are indicated in the GOES light curves in panel (a). We observe that ejecta I precedes the flare while ejecta II occurs during the impulsive phase, shortly before the peak. In panels (d) and (h), we mark a wide circular ribbon structure by sky blue arrows, situated north of the mixed-polarity core region. The compact post-flare loops, formed as a result of standard flare reconnection between the legs of the field lines stretched by the erupting flux rope, are indicated by a red arrow (panel (d)). We explain the formation of the circular ribbon and the compact post-flare loops with the help of a schmatic diagram (Figure \ref{fig:schematic}) in Section \ref{sec:discussion}.
Afterward, we observe a gradual decline in the light curves, indicating the decay phase, which is marked by the growth of dense post-flare loop arcades (shown in panels (e) and (i)). Multiple eruptions in close succession like this have been observed before, and it is plausible that the first eruption leads to a destabilization of nearby fields in the same region leading to the second eruption \citep[e.g.,][]{Torok2011,Sterling2014,Joshi2020}.

In Figure \ref{fig:AIA_M2.6}(a), we show the temporal variation of the second (M2.6 flare) event. In the following panels, we demonstrate the structural changes associated with the eruption's evolution in EUV 94 \AA\ and 171 \AA\ images. During the pre-flare stage, the coronal configuration of the flaring region shows similarity to that of event I in the form of small-scale connectivities and the existence of quasi-open-type field lines (see panels (b) and (f)).
Similar to event I, here also we observe two discrete eruptions in association with the flaring activities. The first eruption starts in the eastern part of the core region just after the beginning of impulsive phase at $\approx$23:46 UT.
We mark this eruption as `ejecta I' (panels (c) and (g)). The second eruption originates from the western part of the core just after the ejecta I at $\approx$23:48 UT, which we term as `ejecta II' (shown in panels (d) and (h)). These successive eruptions show similar morphological behaviour to that of event I with respect to their origin and the subsequent path followed by them. The flare reaches its peak at $\approx$23:51 UT (see panel (a)). Thereafter, the post-flare loops are observed to form (shown in panels (e) and (i)).

\begin{figure}[htp!]

\hspace*{6.5cm}\includegraphics[scale=0.2]{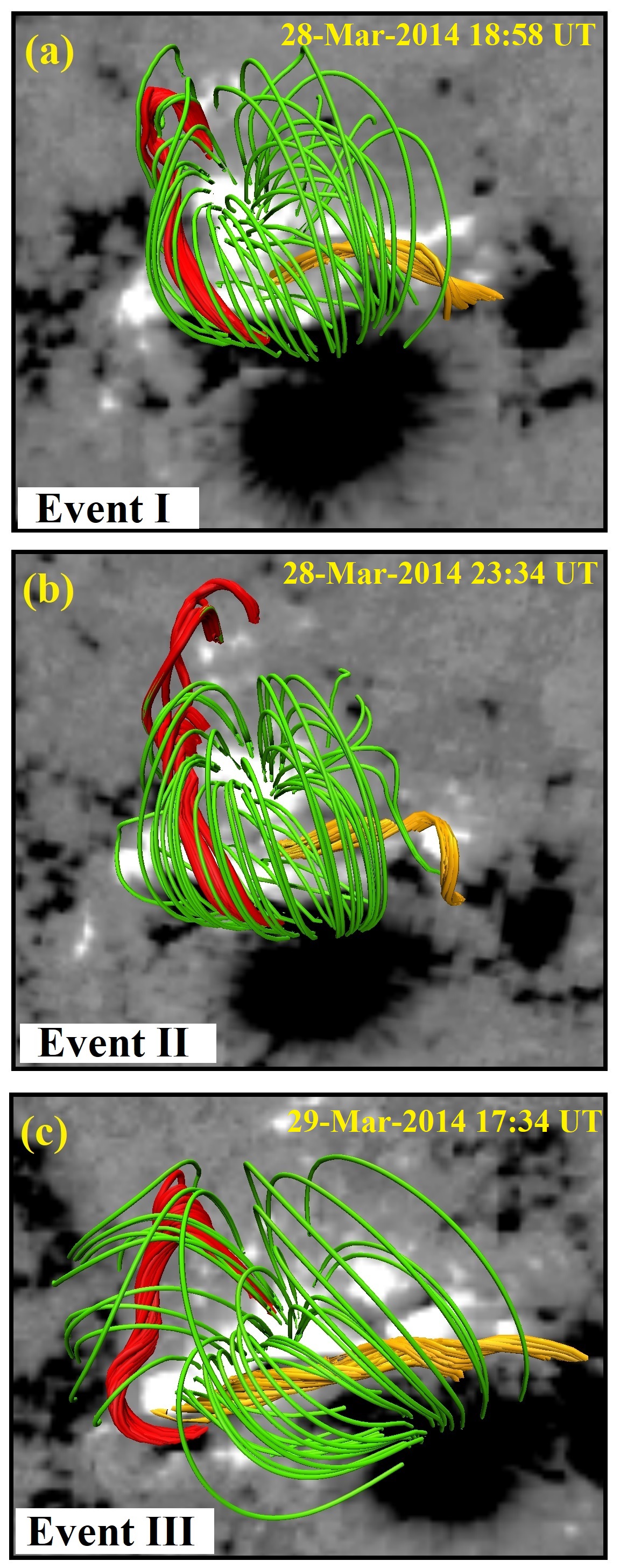}
\caption{Pre-eruptive configurations of the core region in zoomed view, obtained from the NLFFF extrapolations using HMI vector magnetograms. The core consists of closed bipolar field lines (shown by green) constraining the underlying flux ropes. The flux ropes form over the strong mixed polarity region within the core. We show the flux ropes lying over the eastern and western parts of the core region by red and yellow field lines, respectively.}
\label{fig:flux_ropes}
\end{figure}

The temporal and spatial evolutionary phases of the third (X1.0 flare) event are depicted in Figure \ref{fig:AIA_X1.0}. During the pre-eruptive stage of the eruption, we note the existence of a null-point-like structure connecting the opposite magnetic polarities of the flaring region, which is evident in the 171 \AA\ image in panel (g). Unlike the two previous events, both of which consisted of two ejective episodes, in this case there is only a single eruption from the core region, which starts at $\approx$17:45 UT, as marked in panel (a). The eruption apparently destroys the null-point-like structure during the buildup to the maximum phase of the X1.0 flare. The flare peaks at $\approx$17:48 UT (indicated in panel (a)). After that, the dense post-flare arcades are formed (shown in panels (e) and (i)).

Notably, we observe circular ribbon structures during the peak of all the events (see panels (d) and (h) of Figures \ref{fig:AIA_M2.0}, \ref{fig:AIA_M2.6}, and \ref{fig:AIA_X1.0}).

To understand the magnetic complexities of the core region on the size scale of the AR, we employ coronal magnetic field modeling using the NLFFF extrapolation technique. We demonstrate the results of the extrapolation carried out during the pre-flare stages of the events in Figure \ref{fig:flux_ropes}. The extrapolation results clearly demonstrate the existence of two adjacent flux rope system for each of the three events. We note that the compact flux ropes lie over the compact region of strong mixed polarity within the core region (marked by a green box in Figure \ref{fig:pfss1}(b)). The flux ropes lying on the eastern and western parts of the core are shown by red and yellow field lines, respectively. We note a system of low-lying closed field lines (shown in green) connecting the negative and positive polarities of the core that constrain the two flux ropes.
A comparison of modelled coronal field structure (Figure \ref{fig:flux_ropes}) with the corresponding imaging observations (Figures \ref{fig:AIA_M2.0}, \ref{fig:AIA_M2.6}, and \ref{fig:AIA_X1.0}) suggests sequential eruptions of the eastern and western flux ropes during the events I and II, whereas, only a single flux rope erupts for the case of event III. For event III, observational results suggest the eruption of western flux rope (shown by yellow field lines in Figure \ref{fig:flux_ropes}(c)) from the core region.

Even though the third eruption does not have two ``ejecta'' as in the first two events, there is nonetheless enhanced activity in this third event prior to the main eruption. It is visible in the GOES plot of Figure \ref{fig:AIA_X1.0}(a) peaking shortly after 17:40 UT, and it corresponds to an initial movement of the filament prior to eruption, with accompanying brightenings (visible in 94 \AA, 304 \AA, and 171 \AA\ images). The difference for this third event from those first two is that in this case the pre-flare motions and brightenings are along the same main magnetic neutral line (or along the same portion of that neutral line) from which the main eruption occurs, rather than manifesting as an earlier ``ejecta'' event at a different location in the AR. This is similar to the stop-and-start ``slow-rise'' evolution seen in other filament eruption events \citep[e.g.,][]{Sterling2005}. For each of these three cases, the erupting flux ropes act as a ``seed'' toward the formation of large-scale CME structures.

\begin{figure}[htp!]
\hspace*{1.5cm}\includegraphics[scale=1]{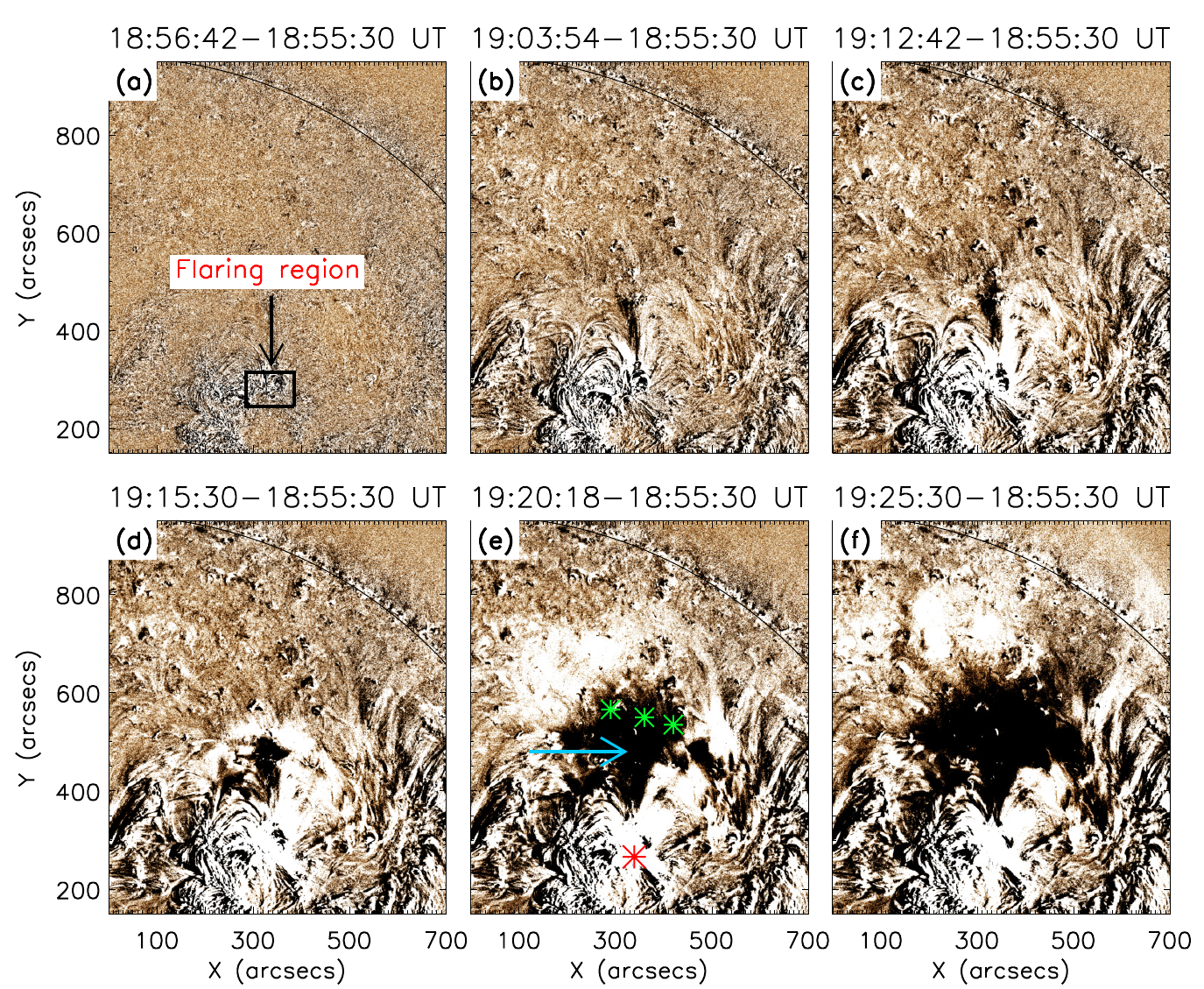}
\caption{Panels (a)--(f): A sequence of AIA 193 \AA\ fixed-difference images showing the coronal dimming accompanying the large-scale eruptions for the case of event I. An image before the start of the flare (at 18:55:30 UT) is subtracted from all the subsequent images. In panel (a), we mark the flaring region by a box. Panel (c) approximately denotes the start of the flare (cf. Figure \ref{fig:AIA_M2.0}(a)). In panel (d), we observe the appearance of slight dimming adjacent to the flaring region. Panel (e) shows the subsequent growth of the dimming, which is marked by an arrow. We indicate the center of the flaring region by a red star and a part of the DPR by green stars. These marked locations denote the footpoints of the large-scale field lines (i.e., magnetic arch) involved in the formation of broad CMEs (see Figures \ref{fig:pfss1}(b)--(c)). Panel (f) shows a later image, in which the dimming has expanded.}

\label{fig:large_scale_M2.0}
\end{figure}

\begin{figure}[htp!]
\hspace*{1.5cm}\includegraphics[scale=1]{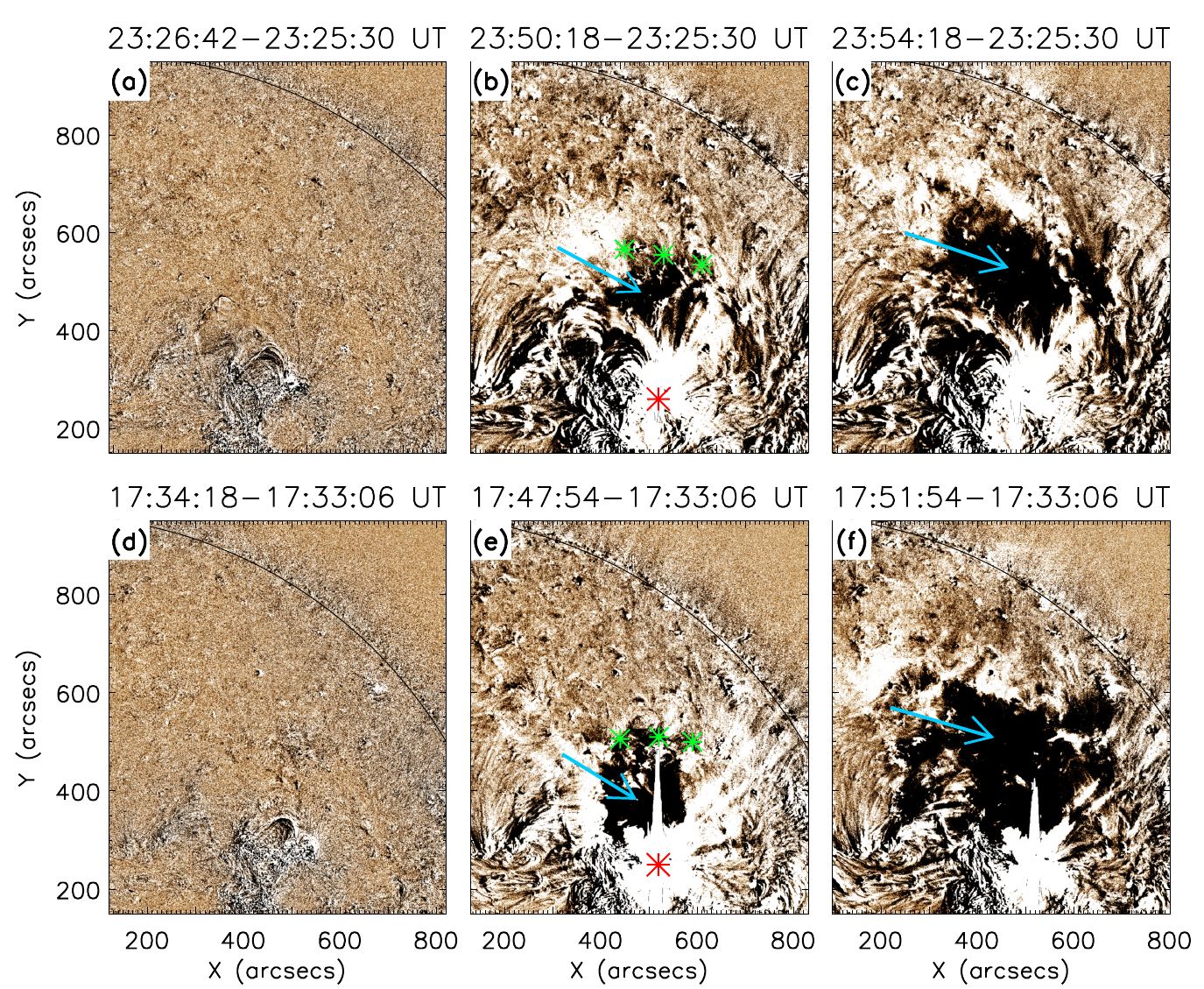}
\caption{Panels (a)--(c): The eruptions from the core region and subsequent appearance of coronal dimming for event II, observed in 193 \AA\ fixed-difference images. The dimmings are indicated in panels (b)--(c) by arrows. Panels (d)--(f): The post-eruption coronal features are depicted in 193 \AA\ fixed-difference images for event III. The dimmings are indicated in panels (e)--(f) by arrows. In panels (b) and (e), we mark the center of the flaring region by red stars and a part of the DPR by green stars. These marked locations are essentially the footpoints of the large-scale field lines (i.e., magnetic arch), whose blowout-eruption resulted in the formation of broad CMEs.}
\label{fig:large_scale_M2.6_X1.0}
\end{figure}

\subsection{Magnetic-arch-blowout and coronal dimming}
\label{sec:large_scale}

All the events analyzed in this study are eruptive in nature, and each of the three eruptions led to the formation of broad CMEs.
Although the CMEs possessed large-scale structures with wide angular width ($>$100$^{\circ}$ to halo; Figure \ref{fig:CME}), the corresponding source active regions of flare blowout-eruptions were much compact (see the spatial comparison shown in Figure \ref{fig:pfss1}(b)). This phenomenon of the CME being much wider than the source region has been recognized for some time \citep[e.g.,][]{Harrison1995,Dere1997,Gopalswamy2000}. \citet{Moore2007a} argue that such a widening between the source region and the CME is a consequence of the magnetic pressure of the exploding field coming into pressure balance with the interplanetary field in the solar wind, which is far weaker than the AR coronal field surrounding the source region, meaning that the CME field has to expand substantially for that new pressure balance to ensue.
We show with AIA 193 \AA\ fixed-difference images the large-scale coronal changes accompanying the early evolution and subsequent development of the broad CMEs during our three events.
Some snapshots of these observations during the course of event I are represented in Figure \ref{fig:large_scale_M2.0}. 
Note that the FOV chosen in Figures \ref{fig:large_scale_M2.0}(a)--(f) represents a much larger area compared to the FOV shown in Figures \ref{fig:AIA_M2.0}, \ref{fig:AIA_M2.6}, and \ref{fig:AIA_X1.0}. For comparison, in Figure \ref{fig:large_scale_M2.0}(a), we indicate the flaring region (cf. sky blue box in Figure \ref{fig:pfss1}(b)).
The saturated pixels in and around the flaring region in panel (c) approximately mark the start of the flare. 
The eruption of flux ropes from the core region is followed by EUV dimming. The onset of the dimming can be realized in the form of a slight dark region adjacent to the flaring region which is manifested as a result of sudden plasma depletion (see Figure \ref{fig:large_scale_M2.0}(d)).
The dimming region expanded gradually which is indicated by an arrow in Figure \ref{fig:large_scale_M2.0}(e), spreading out as an ``EIT Wave'' (or ``EUV Wave'') \citep[e.g.,][]{Thompson2009,Gallagher2011,Long2014}. In the following panel (f), we show a widespread dimming formed northward of the AR.


The morphological features observed during the course of the eruptions for events II and III are shown in the upper and lower panels of Figure \ref{fig:large_scale_M2.6_X1.0}, respectively. 
Following the flux rope eruptions, the EUV coronal dimming is observed subsequently to grow to cover a large area (shown by sky blue arrows in Figures \ref{fig:large_scale_M2.6_X1.0}(b)--(c)). 

Event III presents much more pronounced large-scale structures compared to events I and II. However, it shows morphological similarities with the previous events in terms of development of coronal dimming and resulting broad CME, which actually becomes a halo CME for event III.
The dimmings are indicated by sky blue arrows in Figures \ref{fig:large_scale_M2.6_X1.0}(e)--(f). 
In Figures \ref{fig:large_scale_M2.0}(e), \ref{fig:large_scale_M2.6_X1.0}(b), and \ref{fig:large_scale_M2.6_X1.0}(e), we mark the location of the flaring region and a part of the DPR by red and green stars, respectively. These locations actually denote the footpoints of the large-scale field lines (i.e., magnetic arch), whose blowout-eruption resulted in the formation of the broad CMEs and accompanying EUV dimming.

\newpage

\section{Discussion}
\label{sec:discussion}

In this study, we explored the formation process of three homologous, broad CMEs resulting from eruptive flares in the compact bipolar region of AR NOAA 12017 over 2014 March 28--29. All the events were identified as flux rope eruptions, formed over the same PIL of the AR. Our work presents a clear example of a large-scale coronal magnetic configuration that is repeatedly blown out by compact flux rope eruptions leading to a series of broad CMEs.

The EUV imaging observations of AR NOAA 12017 clearly reveal filaments at the core of the AR near the polarity inversion lines. The magnetic field holding the filaments along the PIL erupts successively three times within a time span of $\approx$24 hr. Our NLFFF extrapolation results reveal the existence of twisted magnetic structures that would envelope the filaments and are capable of storing the magnetic free energy \citep[e.g.,][]{Fan2007,Zhang2012,Toriumi2019,Mitra2020,Sahu2020} required for the subsequent multiple eruptions. For events I and II, the flux rope containing the filament near the eastern part of the core erupted first (obtained in modeling as the red flux rope structure in Figures \ref{fig:flux_ropes}(a)--(b)) followed by the eruption of flux rope containing the filament from the western part (indicated by yellow flux rope structure in Figures \ref{fig:flux_ropes}(a)--(b)). For event III, we observe a single filament/flux rope eruption from the western part of the core region; the extrapolation results indicate that this is likely triggered by the eruption of the yellow flux rope shown in Figure \ref{fig:flux_ropes}(c). The sequential eruption of filaments led to homologous flares followed by CMEs. The eruptions occurred from a very compact site (i.e, core) within the flaring region (see Figure \ref{fig:pfss1}(b)), while the resulting CMEs are of wide angular width ($>$100$^{\circ}$ to halo) (see Figure \ref{fig:CME}). \citet{Woods2018} explored the triggering mechanism of the filament eruption that occurred with the X-class flare of our study (event III). The authors confirmed the existence of two flux ropes present within the active region prior to flaring. Interestingly, only one of these two flux ropes erupts during the flare. \citet{Woods2018} interpreted that tether-cutting reconnection allowed one of the flux ropes to rise to a torus-unstable region prior to flaring, resulting in its successful eruption.

For exploring the large-scale coronal magnetic field changes causing broad CMEs, we conduct PFSS extrapolation to visualize global potential coronal loops in and around the active region (shown in Figures \ref{fig:pfss1}(a), \ref{fig:pfss1}(c) and \ref{fig:pfss2}(b)--(c)). We observe large-scale field lines connecting the distant positive polarity region (DPR) with the flaring region. In view of the generation of broad CMEs, we propose a scenario in which the erupting magnetic flux ropes disrupt these large-scale coronal loops to evolve into broad CMEs. We note that the open field lines (shown by pink) originating from the flaring region can act as a ``runway'' for the successful successive eruption of the flux ropes. The influence of large-scale open field lines in the kinematic and dynamic evolution of CMEs has also been investigated in some other recent studies \citep[e.g.,][]{Chen2013,Georgoulis2019,Gou2019}.

We examine the AIA 193 \AA\ fixed-difference images during the course of the events over an extended neighbourhood of the flaring region. We observe that the eruptions are followed by substantial coronal dimming \citep{Sterling1997,Reinard2008,Mason2014} which gradually expanded (see Figures \ref{fig:large_scale_M2.0} and \ref{fig:large_scale_M2.6_X1.0}). Previous studies have shown that coronal dimming corresponds to the temporary regions of strongly reduced coronal emission in EUV and soft X-rays that form in the wake of CMEs. In general, their formation is interpreted as density depletion due to the expansion and expulsion of plasma during the early CME evolution. The presence of large-scale open field lines, as demonstrated in the present study, would further support the growth of dimming regions as open field lines act as conduits for outward plasma flow.


\begin{figure}[htp!]
\begin{center}

\includegraphics[scale=0.15]{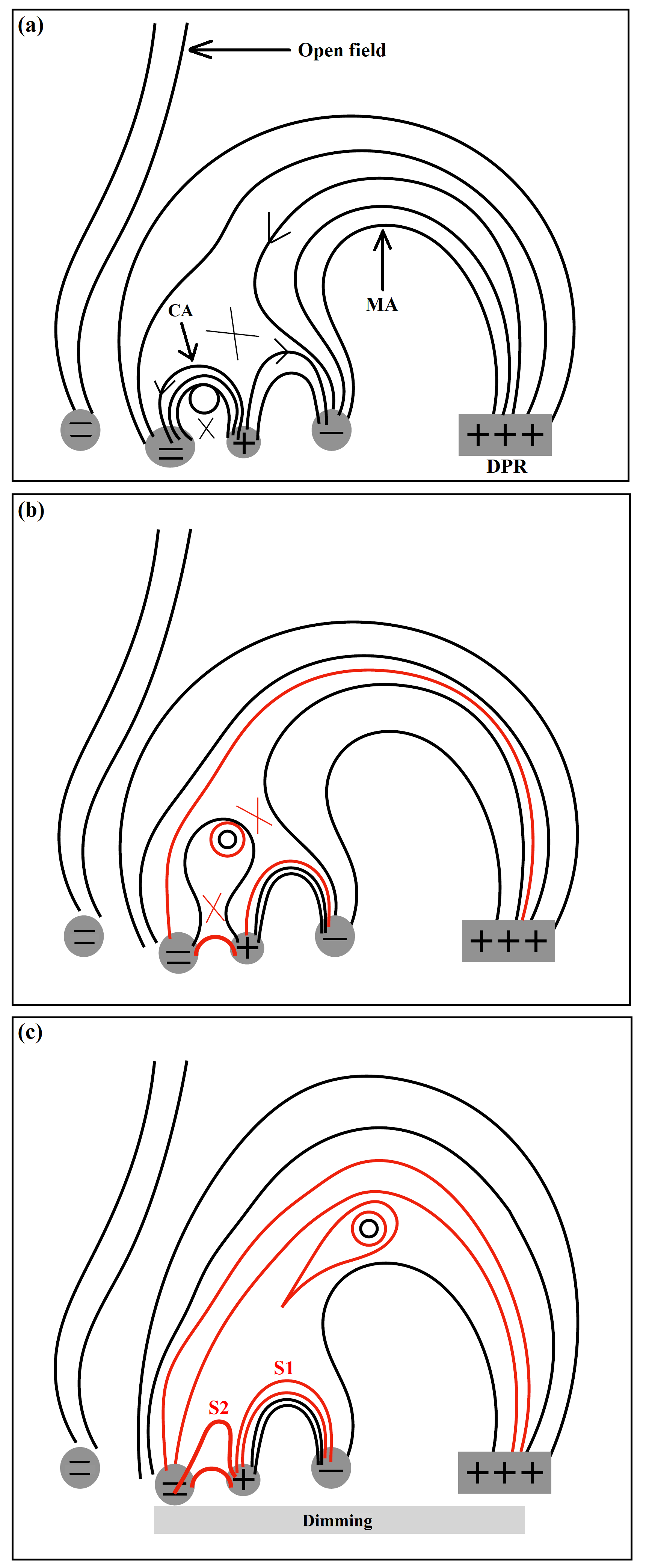}
\caption{Schematic representation of the MAB scenario for the production of broad CMEs resulting from homologous compact major blowout eruptions, viewed from solar west. Panel (a): The large magnetic arch (MA) connects the DPR and the negative flux region of the AR. The large negative sunspot of the AR is denoted by circles with double negative signs, from where the open field lines also originate. On the right of the large sunspot, we show a compact bipolar region hosting the compact arcade (CA), enveloping the flux rope. On the right of CA, we show another set of field lines, which connect the compact region and a larger negative flux region, situated north of the compact region (see panels (b) and (f) of Figures \ref{fig:AIA_M2.0}--\ref{fig:AIA_X1.0}). The plausible reconnection sites are shown by cross signs. Panel (b): The reconnection betweeen CA and MA, weakens the MA field lines and creates a pathway for the eruption of the flux rope. Panel (c): The external reconnection between CA and MA produces the set of field lines labeled S1, and the internal reconnection between the legs of the field lines stretched by the erupting flux rope creates the set of field lines labeled S2. The brightening in the outer footpoints of the S1 field lines are observationally confirmed by the wide circular ribbon structure, whereas, the S2 field lines exhibit themselves as compact post-flare loop arcades (see panels (d) and (h) of Figures \ref{fig:AIA_M2.0}--\ref{fig:AIA_X1.0}). The extent of the dimming resulted from the blowout-eruption of MA is also indicated in panel (c).}  
\label{fig:schematic}
\end{center}
\end{figure}

The blowout-eruptions of compact flux ropes from the core region and their sequential interactions with the overlying large-scale coronal field resulted in the broad CMEs. The present observations exhibit excellent conformity with the MAB scenario originally proposed in \citet{Moore2007}. In essence, the erupting compact flux ropes explode up the large-scale field lines connecting the flaring region of the AR and the DPR (see Figures \ref{fig:pfss1}(b)--(c)). In a feedback process, the activated flux rope blows out the large-scale field lines which in turn strengthens the magnetic field of the erupting CME-flux rope.

In Figure \ref{fig:schematic}, we show a schematic representation (viewed from solar west) of the MAB scenario for the production of the broad CMEs resulting from our three homologous compact major blowout-eruption solar flares. In Figure \ref{fig:schematic}(a), we show the large-scale field (cf. white field lines in Figures \ref{fig:pfss1}(c) and \ref{fig:pfss2}(b)--(c)), which has one end rooted in the DPR, while the other end terminates at a part of the large leading negative sunspot of the AR and at a negative flux region situated north of the compact mixed-polarity region (labeled as `core' in Figure \ref{fig:pfss1}(b)). The large leading negative sunspot of the AR is denoted by circles with double negative signs. The large field lines connecting the DPR and the AR essentially form a `magnetic arch' (MA). The open field lines (cf. pink field lines in Figures \ref{fig:pfss1}(a), (c), and \ref{fig:pfss2}(b)--(c)) originate from the large negative sunspot in the adjacent neighborhood of MA. In Figure \ref{fig:schematic}(a), on the right of the large negative sunspot, we show the positive polarity of the compact region, which hosts the compact arcade (CA; cf. green field lines in Figure \ref{fig:flux_ropes}) enveloping the flux rope. On the right of CA we show another set of field lines; these connect to a distant dispersed negative polarity region, situated north of the compact mixed polarity region (see panels (b) and (f) of Figures \ref{fig:AIA_M2.0}--\ref{fig:AIA_X1.0}). The eruption of the flux rope induces reconnection (i.e., external reconnection) between the CA and MA field lines. Another reconnection, which is standard flare reconnection (i.e., internal reconnetion) will ensue between the legs of the field lines stretched by the erupting flux rope. The plausible reconnection sites are shown by cross signs (Figures \ref{fig:schematic}(a)--(b)) and the post-reconnection loops are drawn in red in Figures \ref{fig:schematic}(b)--(c). The reconnection weakens the MA field lines gradually and creates a ``pathway'' for the subsequent eruption of the flux rope. The eruption of the flux rope continues along the curve of the large MA loops. As this process continues, the flux rope eventually blows out the large MA loops, making the strong dimming region (indicated in figure \ref{fig:schematic}(c)) extending from the AR up to the DPR (observed to form northward of the AR; see Figures \ref{fig:large_scale_M2.0}--\ref{fig:large_scale_M2.6_X1.0}). The extent of the dimming region demarcates the lateral section of the MA that gets blown out and results in the broad CME structure. In Figure \ref{fig:schematic}(c), we indicate two sets (S1 and S2) of post-reconnection field lines, where the observations indicate that S1 is larger in size than S2, since the negative footpoint of S1 connects to a relatively distant region compared to S2 (see panels (b) and (f) of Figures \ref{fig:AIA_M2.0}--\ref{fig:AIA_X1.0}). The brightness along the outer footpoints of S1 field lines is observationally confirmed by a wide circular ribbon structure, that is clearly visible during the peak of the flares (see panels (d) and (h) of Figures \ref{fig:AIA_M2.0}--\ref{fig:AIA_X1.0}). The S2 field lines exhibit themselves as relatively compact post-flare loop arcades (indicated by the red arrow in Figure \ref{fig:AIA_M2.0}(d); see also Figure \ref{fig:AIA_M2.0}(h), panels (d) and (h) of Figures \ref{fig:AIA_M2.6}--\ref{fig:AIA_X1.0}), appearing within the circular ribbon periphery.  
As the eruption of the flux rope continues, it is channeled along the MA structure.
Thus, the flux rope experiences substantial deviation from its original path, as the eruption proceeds. This kind of lateral deflection during the eruption was also observed in previous studies of `over-and-out' type CMEs \citep{Jiang2009,Yang2011,Yang2012a,Yang2012}. We observe large surges from the flaring region after the eruption of the flux ropes in all the events. The surges, consisting of cool plasma expelled from the flaring region, nearly follow the open field lines shown in Figures \ref{fig:pfss1}(c), \ref{fig:pfss2}, and \ref{fig:schematic}. Notably, a significant portion of the surge erupted from the eastern part of the flaring region for event I and from the western part for the subsequent events, which is likely due to the changes in the magnetic configurations of the flaring region. 

Between our study and the study of \citet{Moore2007}, there are some similarities as well as dissimilarities in terms of the pre-eruptive configuration of different observational features detected, but in both cases the central idea involving the physics of eruption remains the same. Unlike the eruption in \citet{Moore2007}, the eruptions in our analysis do not occur in the foot of one leg of a large magnetic arch in the base of a large streamer; another difference is that our CMEs have greater angular widths than did the \citet{Moore2007} CME. On the other hand, our case and the \citet{Moore2007} paper have critical similarities: e.g., both studies have compact ejective flares seated at one foot of a large magnetic arch, and in both cases the origin of the CMEs occurred laterally far offset from the flare site. In view of this, we note that our analysis presents important observational evidence of the MAB scenario for CME formation resulting from homologous compact major blowout-eruption solar flares. Our work essentially generalizes the MAB mechanism formulated in \citet{Moore2007} to more general cases, including to cases with homologous flares.

To summarize, our study incorporates a comprehensive analysis of three homologous ejective eruptive events triggered by a sequence of three compact flux rope eruptions and subsequent blowout of three broad CMEs. The eruptions produce flares of successively increasing intensities (M2.0, M2.6, and X1.0), and generate large scale EUV dimmings. 
The occurrence of homologous and broad CMEs has important consequences for space weather conditions. A comprehensive understanding of such events and their generation mechanism is vital as the space age progresses.

\acknowledgements

A.C.S. and R.L.M. acknowledge support from the NASA HGI and HSR programs. The authors would like to thank the SDO and SOHO teams for their open data policy. SDO (2010--present) is NASA's mission under the Living With a Star (LWS) program. SOHO (1995--present) is a joint project of international cooperation between the ESA and NASA. We thank the anonymous referee for providing insightful comments and suggestions, which improved the presentation and overall quality of the article.

\software{SolarSoftWare \citep{Freeland2012}, VAPOR \citep{Clyne2007}.}

\bibliography{ms.bib}{}

\begin{thebibliography}{}
\expandafter\ifx\csname natexlab\endcsname\relax\def\natexlab#1{#1}\fi

\bibitem[{{Asai} {et~al.}(2003){Asai}, {Ishii}, {Kurokawa}, {Yokoyama}, \&
  {Shimojo}}]{Asai2003}
{Asai}, A., {Ishii}, T.~T., {Kurokawa}, H., {Yokoyama}, T., \& {Shimojo}, M.
  2003, \apj, 586, 624

\bibitem[{{Aulanier} {et~al.}(2012){Aulanier}, {Janvier}, \&
  {Schmieder}}]{Aulanier2012}
{Aulanier}, G., {Janvier}, M., \& {Schmieder}, B. 2012, \aap, 543, A110

\bibitem[{{Aulanier} {et~al.}(2005){Aulanier}, {Pariat}, \&
  {D{\'e}moulin}}]{Aulanier2005}
{Aulanier}, G., {Pariat}, E., \& {D{\'e}moulin}, P. 2005, \aap, 444, 961

\bibitem[{{Aulanier} {et~al.}(2006){Aulanier}, {Pariat}, {D{\'e}moulin}, \&
  {DeVore}}]{Aulanier2006}
{Aulanier}, G., {Pariat}, E., {D{\'e}moulin}, P., \& {DeVore}, C.~R. 2006,
  \solphys, 238, 347

\bibitem[{{Aulanier} {et~al.}(2010){Aulanier}, {T{\"o}r{\"o}k}, {D{\'e}moulin},
  \& {DeLuca}}]{Aulanier2010}
{Aulanier}, G., {T{\"o}r{\"o}k}, T., {D{\'e}moulin}, P., \& {DeLuca}, E.~E.
  2010, \apj, 708, 314

\bibitem[{{Bemporad} {et~al.}(2005){Bemporad}, {Sterling}, {Moore}, \&
  {Poletto}}]{Bemporad2005}
{Bemporad}, A., {Sterling}, A.~C., {Moore}, R.~L., \& {Poletto}, G. 2005,
  \apjl, 635, L189

\bibitem[{{Benz}(2008)}]{Benz2008}
{Benz}, A.~O. 2008, Living Reviews in Solar Physics, 5, 1

\bibitem[{{Brueckner} {et~al.}(1995){Brueckner}, {Howard}, {Koomen},
  {Korendyke}, {Michels}, {Moses}, {Socker}, {Dere}, {Lamy}, {Llebaria},
  {Bout}, {Schwenn}, {Simnett}, {Bedford}, \& {Eyles}}]{Brueckner1995}
{Brueckner}, G.~E., {Howard}, R.~A., {Koomen}, M.~J., {et~al.} 1995, \solphys,
  162, 357

\bibitem[{{Cai} {et~al.}(2021){Cai}, {Zhang}, {Ning}, {Su}, \& {Ji}}]{Cai2021}
{Cai}, Z.~M., {Zhang}, Q.~M., {Ning}, Z.~J., {Su}, Y.~N., \& {Ji}, H.~S. 2021,
  \solphys, 296, 61

\bibitem[{{Carmichael}(1964)}]{Carmichael1964}
{Carmichael}, H. 1964, {A Process for Flares}, Vol.~50, 451

\bibitem[{{Chandra} {et~al.}(2009){Chandra}, {Schmieder}, {Aulanier}, \&
  {Malherbe}}]{Chandra2009}
{Chandra}, R., {Schmieder}, B., {Aulanier}, G., \& {Malherbe}, J.~M. 2009,
  \solphys, 258, 53

\bibitem[{{Chatterjee} \& {Fan}(2013)}]{Chatterjee2013}
{Chatterjee}, P., \& {Fan}, Y. 2013, \apjl, 778, L8

\bibitem[{Chen(2013)}]{Chen2013}
Chen, Y. 2013, Chinese Science Bulletin, 58, doi:10.1007/s11434-013-5669-6

\bibitem[{{Chertok} {et~al.}(2004){Chertok}, {Grechnev}, {Hudson}, \&
  {Nitta}}]{Chertok2004}
{Chertok}, I.~M., {Grechnev}, V.~V., {Hudson}, H.~S., \& {Nitta}, N.~V. 2004,
  Journal of Geophysical Research (Space Physics), 109, A02112

\bibitem[{{Chintzoglou} {et~al.}(2019){Chintzoglou}, {Zhang}, {Cheung}, \&
  {Kazachenko}}]{Chintzoglou2019}
{Chintzoglou}, G., {Zhang}, J., {Cheung}, M. C.~M., \& {Kazachenko}, M. 2019,
  \apj, 871, 67

\bibitem[{{Clyne} {et~al.}(2007){Clyne}, {Mininni}, {Norton}, \&
  {Rast}}]{Clyne2007}
{Clyne}, J., {Mininni}, P., {Norton}, A., \& {Rast}, M. 2007, New Journal of
  Physics, 9, 301

\bibitem[{{Demoulin} {et~al.}(1997){Demoulin}, {Bagala}, {Mandrini}, {Henoux},
  \& {Rovira}}]{Demoulin1997}
{Demoulin}, P., {Bagala}, L.~G., {Mandrini}, C.~H., {Henoux}, J.~C., \&
  {Rovira}, M.~G. 1997, \aap, 325, 305

\bibitem[{{Dere} {et~al.}(1997){Dere}, {Brueckner}, {Howard}, {Koomen},
  {Korendyke}, {Kreplin}, {Michels}, {Moses}, {Moulton}, {Socker}, {St. Cyr},
  {Delaboudini{\`e}re}, {Artzner}, {Brunaud}, {Gabriel}, {Hochedez}, {Millier},
  {Song}, {Chauvineau}, {Marioge}, {Defise}, {Jamar}, {Rochus}, {Catura},
  {Lemen}, {Gurman}, {Neupert}, {Clette}, {Cugnon}, {van Dessel}, {Lamy},
  {Llebaria}, {Schwenn}, \& {Simnett}}]{Dere1997}
{Dere}, K.~P., {Brueckner}, G.~E., {Howard}, R.~A., {et~al.} 1997, \solphys,
  175, 601

\bibitem[{{Devi} {et~al.}(2020){Devi}, {Joshi}, {Chandra}, {Mitra}, {Veronig},
  \& {Joshi}}]{Devi2020}
{Devi}, P., {Joshi}, B., {Chandra}, R., {et~al.} 2020, \solphys, 295, 75

\bibitem[{{Domingo} {et~al.}(1995){Domingo}, {Fleck}, \&
  {Poland}}]{Domingo1995}
{Domingo}, V., {Fleck}, B., \& {Poland}, A.~I. 1995, \ssr, 72, 81

\bibitem[{{Fan} \& {Gibson}(2007)}]{Fan2007}
{Fan}, Y., \& {Gibson}, S.~E. 2007, \apj, 668, 1232

\bibitem[{{Fisk}(2005)}]{Fisk2005}
{Fisk}, L.~A. 2005, \apj, 626, 563

\bibitem[{{Fletcher} \& {Hudson}(2002)}]{Fletcher2002}
{Fletcher}, L., \& {Hudson}, H.~S. 2002, \solphys, 210, 307

\bibitem[{{Freeland} \& {Handy}(2012)}]{Freeland2012}
{Freeland}, S.~L., \& {Handy}, B.~N. 2012, {SolarSoft: Programming and data
  analysis environment for solar physics}, , , ascl:1208.013

\bibitem[{{Gallagher} \& {Long}(2011)}]{Gallagher2011}
{Gallagher}, P.~T., \& {Long}, D.~M. 2011, \ssr, 158, 365

\bibitem[{{Georgoulis} {et~al.}(2019){Georgoulis}, {Nindos}, \&
  {Zhang}}]{Georgoulis2019}
{Georgoulis}, M.~K., {Nindos}, A., \& {Zhang}, H. 2019, Philosophical
  Transactions of the Royal Society of London Series A, 377, 20180094

\bibitem[{{Gopalswamy} \& {Thompson}(2000)}]{Gopalswamy2000}
{Gopalswamy}, N., \& {Thompson}, B.~J. 2000, Journal of Atmospheric and
  Solar-Terrestrial Physics, 62, 1457

\bibitem[{{Gou} {et~al.}(2019){Gou}, {Liu}, {Kliem}, {Wang}, \&
  {Veronig}}]{Gou2019}
{Gou}, T., {Liu}, R., {Kliem}, B., {Wang}, Y., \& {Veronig}, A.~M. 2019,
  Science Advances, 5, 7004

\bibitem[{{Green} {et~al.}(2011){Green}, {Kliem}, \& {Wallace}}]{Green2011}
{Green}, L.~M., {Kliem}, B., \& {Wallace}, A.~J. 2011, \aap, 526, A2

\bibitem[{{Green} {et~al.}(2002){Green}, {Matthews}, {van Driel-Gesztelyi},
  {Harra}, \& {Culhane}}]{Green2002}
{Green}, L.~M., {Matthews}, S.~A., {van Driel-Gesztelyi}, L., {Harra}, L.~K.,
  \& {Culhane}, J.~L. 2002, \solphys, 205, 325

\bibitem[{{Green} {et~al.}(2018){Green}, {T{\"o}r{\"o}k}, {Vr{\v{s}}nak},
  {Manchester}, \& {Veronig}}]{Green2018}
{Green}, L.~M., {T{\"o}r{\"o}k}, T., {Vr{\v{s}}nak}, B., {Manchester}, W., \&
  {Veronig}, A. 2018, \ssr, 214, 46

\bibitem[{{Harrison}(1995)}]{Harrison1995}
{Harrison}, R.~A. 1995, \aap, 304, 585

\bibitem[{{Hirayama}(1974)}]{Hirayama1974}
{Hirayama}, T. 1974, \solphys, 34, 323

\bibitem[{{Janvier} {et~al.}(2014){Janvier}, {Aulanier}, {Bommier},
  {Schmieder}, {D{\'e}moulin}, \& {Pariat}}]{Janvier2014}
{Janvier}, M., {Aulanier}, G., {Bommier}, V., {et~al.} 2014, \apj, 788, 60

\bibitem[{{Jiang} {et~al.}(2021){Jiang}, {Feng}, {Liu}, {Yan}, {Hu}, {Moore},
  {Duan}, {Cui}, {Zuo}, {Wang}, \& {Wei}}]{Jiang2021}
{Jiang}, C., {Feng}, X., {Liu}, R., {et~al.} 2021, Nature Astronomy,
  doi:10.1038/s41550-021-01414-z

\bibitem[{{Jiang} {et~al.}(2009){Jiang}, {Yang}, {Zheng}, {Bi}, \&
  {Yang}}]{Jiang2009}
{Jiang}, Y., {Yang}, J., {Zheng}, R., {Bi}, Y., \& {Yang}, X. 2009, \apj, 693,
  1851

\bibitem[{{Jing} {et~al.}(2008){Jing}, {Chae}, \& {Wang}}]{Jing2008}
{Jing}, J., {Chae}, J., \& {Wang}, H. 2008, \apjl, 672, L73

\bibitem[{{Joshi} {et~al.}(2018){Joshi}, {Ibrahim}, {Shanmugaraju}, \&
  {Chakrabarty}}]{Joshi2018}
{Joshi}, B., {Ibrahim}, M.~S., {Shanmugaraju}, A., \& {Chakrabarty}, D. 2018,
  \solphys, 293, 107

\bibitem[{{Joshi} {et~al.}(2017{\natexlab{a}}){Joshi}, {Kushwaha}, {Veronig},
  {Dhara}, {Shanmugaraju}, \& {Moon}}]{Joshi2017}
{Joshi}, B., {Kushwaha}, U., {Veronig}, A.~M., {et~al.} 2017{\natexlab{a}},
  \apj, 834, 42

\bibitem[{{Joshi} {et~al.}(2009){Joshi}, {Veronig}, {Cho}, {Bong}, {Somov},
  {Moon}, {Lee}, {Manoharan}, \& {Kim}}]{Joshi2009}
{Joshi}, B., {Veronig}, A., {Cho}, K.~S., {et~al.} 2009, \apj, 706, 1438

\bibitem[{{Joshi} {et~al.}(2021){Joshi}, {Joshi}, \& {Mitra}}]{Joshi2021}
{Joshi}, N.~C., {Joshi}, B., \& {Mitra}, P.~K. 2021, \mnras, 501, 4703

\bibitem[{{Joshi} {et~al.}(2020){Joshi}, {Sterling}, {Moore}, \&
  {Joshi}}]{Joshi2020}
{Joshi}, N.~C., {Sterling}, A.~C., {Moore}, R.~L., \& {Joshi}, B. 2020, \apj,
  901, 38

\bibitem[{{Joshi} {et~al.}(2017{\natexlab{b}}){Joshi}, {Sterling}, {Moore},
  {Magara}, \& {Moon}}]{NavinJoshi2017}
{Joshi}, N.~C., {Sterling}, A.~C., {Moore}, R.~L., {Magara}, T., \& {Moon},
  Y.-J. 2017{\natexlab{b}}, \apj, 845, 26

\bibitem[{{Kharayat} {et~al.}(2021){Kharayat}, {Joshi}, {Mitra}, {Manoharan},
  \& {Monstein}}]{Kharayat2021}
{Kharayat}, H., {Joshi}, B., {Mitra}, P.~K., {Manoharan}, P.~K., \& {Monstein},
  C. 2021, \solphys, 296, 99

\bibitem[{{Kleint} {et~al.}(2015){Kleint}, {Battaglia}, {Reardon}, {Sainz
  Dalda}, {Young}, \& {Krucker}}]{Kleint2015}
{Kleint}, L., {Battaglia}, M., {Reardon}, K., {et~al.} 2015, \apj, 806, 9

\bibitem[{{Kopp} \& {Pneuman}(1976)}]{Kopp1976}
{Kopp}, R.~A., \& {Pneuman}, G.~W. 1976, \solphys, 50, 85

\bibitem[{{Kushwaha} {et~al.}(2014){Kushwaha}, {Joshi}, {Cho}, {Veronig},
  {Tiwari}, \& {Mathew}}]{Kushwaha2014}
{Kushwaha}, U., {Joshi}, B., {Cho}, K.-S., {et~al.} 2014, \apj, 791, 23

\bibitem[{{Lemen} {et~al.}(2012){Lemen}, {Title}, {Akin}, {Boerner}, {Chou},
  {Drake}, {Duncan}, {Edwards}, {Friedlaender}, {Heyman}, {Hurlburt}, {Katz},
  {Kushner}, {Levay}, {Lindgren}, {Mathur}, {McFeaters}, {Mitchell}, {Rehse},
  {Schrijver}, {Springer}, {Stern}, {Tarbell}, {Wuelser}, {Wolfson}, {Yanari},
  {Bookbinder}, {Cheimets}, {Caldwell}, {Deluca}, {Gates}, {Golub}, {Park},
  {Podgorski}, {Bush}, {Scherrer}, {Gummin}, {Smith}, {Auker}, {Jerram},
  {Pool}, {Soufli}, {Windt}, {Beardsley}, {Clapp}, {Lang}, \&
  {Waltham}}]{Lemen2012}
{Lemen}, J.~R., {Title}, A.~M., {Akin}, D.~J., {et~al.} 2012, \solphys, 275, 17

\bibitem[{{Li} {et~al.}(2015){Li}, {Ding}, {Qiu}, \& {Cheng}}]{Li2015}
{Li}, Y., {Ding}, M.~D., {Qiu}, J., \& {Cheng}, J.~X. 2015, \apj, 811, 7

\bibitem[{{Liu} {et~al.}(2015){Liu}, {Deng}, {Liu}, {Lee}, {Pariat},
  {Wiegelmann}, {Liu}, {Kleint}, \& {Wang}}]{Liu2015}
{Liu}, C., {Deng}, N., {Liu}, R., {et~al.} 2015, \apjl, 812, L19

\bibitem[{{Liu} {et~al.}(2017){Liu}, {Wang}, {Liu}, {Zhou}, {Temmer},
  {Thalmann}, {Liu}, {Liu}, {Shen}, {Zhang}, \& {Veronig}}]{Liu2017}
{Liu}, L., {Wang}, Y., {Liu}, R., {et~al.} 2017, \apj, 844, 141

\bibitem[{{Long} {et~al.}(2014){Long}, {Bloomfield}, {Gallagher}, \&
  {P{\'e}rez-Su{\'a}rez}}]{Long2014}
{Long}, D.~M., {Bloomfield}, D.~S., {Gallagher}, P.~T., \&
  {P{\'e}rez-Su{\'a}rez}, D. 2014, \solphys, 289, 3279

\bibitem[{{Longcope}(2005)}]{Longcope2005}
{Longcope}, D.~W. 2005, Living Reviews in Solar Physics, 2, 7

\bibitem[{{Mason} {et~al.}(2014){Mason}, {Woods}, {Caspi}, {Thompson}, \&
  {Hock}}]{Mason2014}
{Mason}, J.~P., {Woods}, T.~N., {Caspi}, A., {Thompson}, B.~J., \& {Hock},
  R.~A. 2014, \apj, 789, 61

\bibitem[{{Masson} {et~al.}(2012){Masson}, {Aulanier}, {Pariat}, \&
  {Klein}}]{Masson2012}
{Masson}, S., {Aulanier}, G., {Pariat}, E., \& {Klein}, K.~L. 2012, \solphys,
  276, 199

\bibitem[{{Masson} {et~al.}(2009){Masson}, {Pariat}, {Aulanier}, \&
  {Schrijver}}]{Masson2009}
{Masson}, S., {Pariat}, E., {Aulanier}, G., \& {Schrijver}, C.~J. 2009, \apj,
  700, 559

\bibitem[{{Mitra} \& {Joshi}(2021)}]{Mitra2021}
{Mitra}, P.~K., \& {Joshi}, B. 2021, \mnras, 503, 1017

\bibitem[{{Mitra} {et~al.}(2020){Mitra}, {Joshi}, \& {Prasad}}]{Mitra2020}
{Mitra}, P.~K., {Joshi}, B., \& {Prasad}, A. 2020, \solphys, 295, 29

\bibitem[{{Mitra} {et~al.}(2018){Mitra}, {Joshi}, {Prasad}, {Veronig}, \&
  {Bhattacharyya}}]{Mitra2018}
{Mitra}, P.~K., {Joshi}, B., {Prasad}, A., {Veronig}, A.~M., \&
  {Bhattacharyya}, R. 2018, \apj, 869, 69

\bibitem[{{Moore} \& {Sterling}(2007)}]{Moore2007}
{Moore}, R.~L., \& {Sterling}, A.~C. 2007, \apj, 661, 543

\bibitem[{{Moore} {et~al.}(2001){Moore}, {Sterling}, {Hudson}, \&
  {Lemen}}]{Moore2001}
{Moore}, R.~L., {Sterling}, A.~C., {Hudson}, H.~S., \& {Lemen}, J.~R. 2001,
  \apj, 552, 833

\bibitem[{{Moore} {et~al.}(2007){Moore}, {Sterling}, \& {Suess}}]{Moore2007a}
{Moore}, R.~L., {Sterling}, A.~C., \& {Suess}, S.~T. 2007, \apj, 668, 1221

\bibitem[{{Owens} {et~al.}(2020){Owens}, {Lockwood}, {Macneil}, \&
  {Stansby}}]{Owens2020}
{Owens}, M., {Lockwood}, M., {Macneil}, A., \& {Stansby}, D. 2020, \solphys,
  295, 37

\bibitem[{{Pallavicini} {et~al.}(1977){Pallavicini}, {Serio}, \&
  {Vaiana}}]{Pallavicini1977}
{Pallavicini}, R., {Serio}, S., \& {Vaiana}, G.~S. 1977, \apj, 216, 108

\bibitem[{{Pesnell} {et~al.}(2012){Pesnell}, {Thompson}, \&
  {Chamberlin}}]{Pesnell2012}
{Pesnell}, W.~D., {Thompson}, B.~J., \& {Chamberlin}, P.~C. 2012, \solphys,
  275, 3

\bibitem[{{Pontin} {et~al.}(2007){Pontin}, {Bhattacharjee}, \&
  {Galsgaard}}]{Pontin2007a}
{Pontin}, D.~I., {Bhattacharjee}, A., \& {Galsgaard}, K. 2007, Physics of
  Plasmas, 14, 052106

\bibitem[{{Prasad} {et~al.}(2020){Prasad}, {Dissauer}, {Hu}, {Bhattacharyya},
  {Veronig}, {Kumar}, \& {Joshi}}]{Prasad2020}
{Prasad}, A., {Dissauer}, K., {Hu}, Q., {et~al.} 2020, \apj, 903, 129

\bibitem[{{Priest} \& {Pontin}(2009)}]{Priest2009}
{Priest}, E.~R., \& {Pontin}, D.~I. 2009, Physics of Plasmas, 16, 122101

\bibitem[{{Rappazzo} {et~al.}(2012){Rappazzo}, {Matthaeus}, {Ruffolo},
  {Servidio}, \& {Velli}}]{Rappazzo2012}
{Rappazzo}, A.~F., {Matthaeus}, W.~H., {Ruffolo}, D., {Servidio}, S., \&
  {Velli}, M. 2012, \apjl, 758, L14

\bibitem[{{Reinard} \& {Biesecker}(2008)}]{Reinard2008}
{Reinard}, A.~A., \& {Biesecker}, D.~A. 2008, \apj, 674, 576

\bibitem[{{Sahu} {et~al.}(2020){Sahu}, {Joshi}, {Mitra}, {Veronig}, \&
  {Yurchyshyn}}]{Sahu2020}
{Sahu}, S., {Joshi}, B., {Mitra}, P.~K., {Veronig}, A.~M., \& {Yurchyshyn}, V.
  2020, \apj, 897, 157

\bibitem[{{Savcheva} {et~al.}(2015){Savcheva}, {Pariat}, {McKillop},
  {McCauley}, {Hanson}, {Su}, {Werner}, \& {DeLuca}}]{Savcheva2015}
{Savcheva}, A., {Pariat}, E., {McKillop}, S., {et~al.} 2015, \apj, 810, 96

\bibitem[{{Schou} {et~al.}(2012){Schou}, {Scherrer}, {Bush}, {Wachter},
  {Couvidat}, {Rabello-Soares}, {Bogart}, {Hoeksema}, {Liu}, {Duvall}, {Akin},
  {Allard}, {Miles}, {Rairden}, {Shine}, {Tarbell}, {Title}, {Wolfson},
  {Elmore}, {Norton}, \& {Tomczyk}}]{Schou2012}
{Schou}, J., {Scherrer}, P.~H., {Bush}, R.~I., {et~al.} 2012, \solphys, 275,
  229

\bibitem[{{Schrijver} {et~al.}(2011){Schrijver}, {Aulanier}, {Title}, {Pariat},
  \& {Delann{\'e}e}}]{Schrijver2011}
{Schrijver}, C.~J., {Aulanier}, G., {Title}, A.~M., {Pariat}, E., \&
  {Delann{\'e}e}, C. 2011, \apj, 738, 167

\bibitem[{{Schrijver} \& {De Rosa}(2003)}]{Schrijver2003}
{Schrijver}, C.~J., \& {De Rosa}, M.~L. 2003, \solphys, 212, 165

\bibitem[{{Schwenn}(2006)}]{Schwenn2006}
{Schwenn}, R. 2006, Living Reviews in Solar Physics, 3, 2

\bibitem[{{Shibata}(1999)}]{Shibata1999a}
{Shibata}, K. 1999, in Proceedings of the Nobeyama Symposium, ed. T.~S.
  {Bastian}, N.~{Gopalswamy}, \& K.~{Shibasaki}, 381--389

\bibitem[{{Shibata} \& {Magara}(2011)}]{Shibata2011}
{Shibata}, K., \& {Magara}, T. 2011, Living Reviews in Solar Physics, 8, 6

\bibitem[{{Sterling} \& {Hudson}(1997)}]{Sterling1997}
{Sterling}, A.~C., \& {Hudson}, H.~S. 1997, \apjl, 491, L55

\bibitem[{{Sterling} \& {Moore}(2005)}]{Sterling2005}
{Sterling}, A.~C., \& {Moore}, R.~L. 2005, \apj, 630, 1148

\bibitem[{{Sterling} {et~al.}(2014){Sterling}, {Moore}, {Falconer}, \&
  {Knox}}]{Sterling2014}
{Sterling}, A.~C., {Moore}, R.~L., {Falconer}, D.~A., \& {Knox}, J.~M. 2014,
  \apjl, 788, L20

\bibitem[{{Sterling} {et~al.}(2011){Sterling}, {Moore}, \&
  {Harra}}]{Sterling2011}
{Sterling}, A.~C., {Moore}, R.~L., \& {Harra}, L.~K. 2011, \apj, 743, 63

\bibitem[{{Sturrock}(1966)}]{Sturrock1966}
{Sturrock}, P.~A. 1966, \nat, 211, 695

\bibitem[{{Sun} {et~al.}(2013){Sun}, {Hoeksema}, {Liu}, {Aulanier}, {Su},
  {Hannah}, \& {Hock}}]{Sun2013}
{Sun}, X., {Hoeksema}, J.~T., {Liu}, Y., {et~al.} 2013, \apj, 778, 139

\bibitem[{{Svestka} \& {Cliver}(1992)}]{Svestka1992}
{Svestka}, Z., \& {Cliver}, E.~W. 1992, {History and Basic Characteristics of
  Eruptive Flares}, ed. Z.~{Svestka}, B.~V. {Jackson}, \& M.~E. {Machado}, Vol.
  399, 1

\bibitem[{{Temmer} {et~al.}(2010){Temmer}, {Veronig}, {Kontar}, {Krucker}, \&
  {Vr{\v{s}}nak}}]{Temmer2010}
{Temmer}, M., {Veronig}, A.~M., {Kontar}, E.~P., {Krucker}, S., \&
  {Vr{\v{s}}nak}, B. 2010, \apj, 712, 1410

\bibitem[{{Thompson} \& {Myers}(2009)}]{Thompson2009}
{Thompson}, B.~J., \& {Myers}, D.~C. 2009, \apjs, 183, 225

\bibitem[{{Toriumi} \& {Wang}(2019)}]{Toriumi2019}
{Toriumi}, S., \& {Wang}, H. 2019, Living Reviews in Solar Physics, 16, 3

\bibitem[{{T{\"o}r{\"o}k} {et~al.}(2011){T{\"o}r{\"o}k}, {Panasenco}, {Titov},
  {Miki{\'c}}, {Reeves}, {Velli}, {Linker}, \& {De Toma}}]{Torok2011}
{T{\"o}r{\"o}k}, T., {Panasenco}, O., {Titov}, V.~S., {et~al.} 2011, \apjl,
  739, L63

\bibitem[{{Vrsnak}(2003)}]{Vrsnak2003}
{Vrsnak}, B. 2003, {Magnetic 3-D Configurations of Energy Release in Solar
  Flares}, ed. L.~{Klein}, Vol. 612, 28--47

\bibitem[{{Wang} \& {Liu}(2012)}]{Wang2012}
{Wang}, H., \& {Liu}, C. 2012, \apj, 760, 101

\bibitem[{{Wang} {et~al.}(2014){Wang}, {Liu}, {Deng}, {Zeng}, {Xu}, {Jing}, \&
  {Cao}}]{Wang2014}
{Wang}, H., {Liu}, C., {Deng}, N., {et~al.} 2014, \apjl, 781, L23

\bibitem[{{Wang} \& {Zhang}(2007)}]{Wang2007}
{Wang}, Y., \& {Zhang}, J. 2007, \apj, 665, 1428

\bibitem[{{Warren} {et~al.}(2011){Warren}, {O'Brien}, \&
  {Sheeley}}]{Warren2011}
{Warren}, H.~P., {O'Brien}, C.~M., \& {Sheeley}, Neil~R., J. 2011, \apj, 742,
  92

\bibitem[{{Webb} \& {Howard}(2012)}]{Webb2012}
{Webb}, D.~F., \& {Howard}, T.~A. 2012, Living Reviews in Solar Physics, 9, 3

\bibitem[{{Wheatland} {et~al.}(2000){Wheatland}, {Sturrock}, \&
  {Roumeliotis}}]{Wheatland2000}
{Wheatland}, M.~S., {Sturrock}, P.~A., \& {Roumeliotis}, G. 2000, \apj, 540,
  1150

\bibitem[{{Wiegelmann}(2004)}]{Wiegelmann2004}
{Wiegelmann}, T. 2004, \solphys, 219, 87

\bibitem[{{Wiegelmann}(2008)}]{Wiegelmann2008}
---. 2008, Journal of Geophysical Research (Space Physics), 113, A03S02

\bibitem[{{Woods} {et~al.}(2018){Woods}, {Inoue}, {Harra}, {Matthews},
  {Kusano}, \& {Kalmoni}}]{Woods2018}
{Woods}, M.~M., {Inoue}, S., {Harra}, L.~K., {et~al.} 2018, \apj, 860, 163

\bibitem[{{Yang} {et~al.}(2012{\natexlab{a}}){Yang}, {Jiang}, {Bi}, {Li},
  {Hong}, {Yang}, {Zheng}, \& {Yang}}]{Yang2012}
{Yang}, J., {Jiang}, Y., {Bi}, Y., {et~al.} 2012{\natexlab{a}}, \apj, 749, 12

\bibitem[{{Yang} {et~al.}(2012{\natexlab{b}}){Yang}, {Jiang}, {Zheng}, {Bi},
  {Hong}, \& {Yang}}]{Yang2012a}
{Yang}, J., {Jiang}, Y., {Zheng}, R., {et~al.} 2012{\natexlab{b}}, \apj, 745, 9

\bibitem[{{Yang} {et~al.}(2011){Yang}, {Jiang}, {Zheng}, {Hong}, {Bi}, \&
  {Yang}}]{Yang2011}
---. 2011, \solphys, 270, 551

\bibitem[{{Yang} {et~al.}(2016){Yang}, {Guo}, \& {Ding}}]{Yang2016}
{Yang}, K., {Guo}, Y., \& {Ding}, M.~D. 2016, \apj, 824, 148

\bibitem[{{Young} {et~al.}(2015){Young}, {Tian}, \& {Jaeggli}}]{Young2015}
{Young}, P.~R., {Tian}, H., \& {Jaeggli}, S. 2015, \apj, 799, 218

\bibitem[{{Zhang} {et~al.}(2012){Zhang}, {Cheng}, \& {Ding}}]{Zhang2012}
{Zhang}, J., {Cheng}, X., \& {Ding}, M.-D. 2012, Nature Communications, 3, 747

\bibitem[{{Zhang} {et~al.}(2001){Zhang}, {Dere}, {Howard}, {Kundu}, \&
  {White}}]{Zhang2001}
{Zhang}, J., {Dere}, K.~P., {Howard}, R.~A., {Kundu}, M.~R., \& {White}, S.~M.
  2001, \apj, 559, 452

\bibitem[{{Zhang} \& {Wang}(2002)}]{Zhang2002}
{Zhang}, J., \& {Wang}, J. 2002, \apjl, 566, L117

\end{thebibliography}
\bibliographystyle{aasjournal}

\end{document}